\documentclass{lmcs}
\pdfoutput=1

\usepackage{lastpage}
\lmcsdoi{17}{1}{5}
\lmcsheading{}{\pageref{LastPage}}{}{}%
{Mar.~20,~2019}{Jan.~25,~2021}{}


\usepackage[utf8]{inputenc}
\usepackage[T1]{fontenc}

\usepackage{amsmath, amsthm, amssymb}
\usepackage{todonotes}

\usepackage[absolute]{textpos}

\usepackage{comment}

\usepackage{thmtools}
\usepackage{thm-restate}

\usepackage{algpseudocode}

\usepackage{xspace}

\usepackage{xfrac}
\usepackage{stmaryrd}

\newcommand{\rank}[1]{\mathrm{rank}(#1)}

\newcommand{\eqdef}{\stackrel{\text{\tiny def}}=}

\newcommand{\Nat}{\mathbb N}
\newcommand{\Nats}{\Nat}

\newcommand{\mso}{{\sc mso}\xspace}
\newcommand{\cmso}{{\sc cmso}\xspace}
\newcommand{\msoone}{{\sc mso}$_1$\xspace}
\newcommand{\cmsoone}{{\sc cmso}$_1$\xspace}
\newcommand{\msotwo}{{\sc mso}$_2$\xspace}
\newcommand{\cmsotwo}{{\sc cmso}$_2$\xspace}

\newcommand{\set}[1]{\{#1\}}

\newcommand{\modulus}[1]{\mathsf{mod}(#1)}
\newcommand{\dist}{\mathrm{dist}}

\newcommand{\abst}[1]{\llbracket #1 \rrbracket}
\newcommand{\dcw}{\mathrm{dcw}}

\newcommand{\structa}{\mathfrak A}

\def\cqedsymbol{\ifmmode$\lrcorner$\else{\unskip\nobreak\hfil
\penalty50\hskip1em\null\nobreak\hfil$\lrcorner$
\parfillskip=0pt\finalhyphendemerits=0\endgraf}\fi} 

\newcommand{\cqed}{\renewcommand{\qed}{\cqedsymbol}}

\newcounter{ourexamplecounter}

\newcounter{runningcounter}

\renewcommand{\setminus}{-}

\usepackage{graphicx}
\usepackage{tikz}
\usetikzlibrary{arrows}
\usetikzlibrary{shapes.geometric}
\usetikzlibrary{calc}

\renewcommand{\leq}{\leqslant}
\renewcommand{\le}{\leqslant}
\renewcommand{\geq}{\geqslant}

\begin{document}

\title{Definable decompositions for graphs of bounded linear cliquewidth}

\author[Mikołaj Bojańczyk]{Mikołaj Bojańczyk\rsuper{a}}

\author[Martin Grohe]{Martin Grohe\rsuper{b}}

\author[Michał Pilipczuk]{Michał Pilipczuk\rsuper{a}}

\address{\lsuper{a}Institute of Informatics, University of Warsaw, Poland}
\email{bojan@mimuw.edu.pl, michal.pilipczuk@mimuw.edu.pl}

\address{\lsuper{b}RWTH Aachen University, Germany}
\email{grohe@informatik.rwth-aachen.de}

\thanks{An extended abstract of this work has appeared in the proceedings of the 33rd Annual ACM/IEEE Symposium on Logic in Computer Science, LICS 2018. This version contains complete proofs of all the claimed results.
  The work of M. Boja\'n{}czyk and Mi. Pilipczuk was supported by the European
    Research Council (ERC) under the European Union's Horizon 2020 research
    and innovation programme (ERC consolidator grant LIPA, agreement no. 683080). 
Mi.~Pilipczuk was supported by the Foundation for Polish Science via the START stipend programme.}




\begin{abstract}
We prove that for every positive integer $k$, there exists an \msoone-transduction that given a graph of linear cliquewidth at most $k$
outputs, nondeterministically, some cliquewidth decomposition of the graph of width bounded by a function of $k$. A direct corollary of this result is the equivalence of
the notions of \cmsoone-definability and recognizability on graphs of bounded linear cliquewidth. 
\end{abstract}

\maketitle

\section{Introduction}\label{sec:intro}

Hierarchical decompositions of graphs have come to play an increasingly important role in logic, algorithms and many other areas of computer science. The treelike structure they impose often allows one to process the data much more efficiently. The best-known and arguably most important graph decompositions are \emph{tree decompositions}, which play a central role in a research direction at the boundary between logic, graph grammars, and generalizations of automata theory to graphs that was pioneered by Courcelle in the 1990s (see \cite{0030804}). Recently, the first and third author of this paper answered a long standing open question in this area by showing that on graphs of bounded treewidth, the automata-theory-inspired notion of \emph{recognizability} coincides with \emph{definability} in monadic second order logic with modulo counting~\cite{bojanczyk2016definability}.

A drawback of tree decompositions is that they only yield meaningful results for sparse graphs. A suitable form of decomposition that also applies to dense graphs 
and that has a similarly nice, yet less developed Courcelle-style theory, is that of \emph{cliquewidth decompositions}, introduced by Courcelle and Olariu~\cite{CourcelleO00}. A cliquewidth decomposition of a graph is a term in a suitable algebra consisting of (roughly) the following operations for constructing and manipulating colored graphs: (i) disjoint union; (ii) for a pair of colors $i,j$,  simultaneously add  an edge for every pair ($i$-colored vertex, $j$-colored vertex);  and (iii) apply a recoloring --- a functional transformation of colors --- to all the vertices. A natural notion of \emph{width} for such a cliquewidth decomposition is the  total number of colors used.
The \emph{cliquewidth} of a graph is the smallest width of a cliquewidth decomposition for it. An alternative notion of graph decomposition that also works well for dense graphs is that of \emph{rank decompositions} introduced by Oum and Seymour \cite{Oum05,OumS06}. The corresponding notion of \emph{rankwidth} turned out to be functionally equivalent to cliquewidth, that is, bounded cliquewidth is the same as bounded rankwidth.
Bounded width clique or rank decompositions may be viewed as hierarchical decompositions that minimize ``modular complexity'' of cuts
present in the decomposition, in the same way as treewidth corresponds to hierarchical decompositions using vertex cuts, where the complexity of a cut is its size.

Cliquewidth is tightly connected to  \msoone logic on graphs, in the same way as the \msotwo logic is connected to treewidth.
Recall that in  \msoone, one can quantify over vertices and sets of vertices, and check their adjacency, while  \msotwo also allows quantification over
 sets of edges.
Both  these logics can be viewed as  plain \mso logic on two different encodings of graphs as relational structures: for \msoone the  encoding uses only vertices as the universe and has a binary adjacency relation, while for \msotwo the encoding uses both vertices and edges as the universe and has an incidence relation binding every edge with its endpoints.
These two logics are connected to cliquewidth and treewidth as follows. 
If a graph property $\Pi$ is definable in \msotwo, then  tree decompositions of graphs in $\Pi$ can be recognized by a finite state device (tree automaton).  This leads, for instance, to a fixed-parameter model checking algorithm for \msotwo-definable
properties on graphs of bounded treewidth~\cite{Courcelle90}.
This notion of recognizability, where tree decompositions are processed, is called {\em{HR-recognizability}}~\cite{0030804}.
Similarly, if $\Pi$ is  \msoone-definable, then cliquewidth decompositions of graphs in $\Pi$ can be recognized by a finite state device. This notion of recognizability is called {\em{VR-recognizability}}~\cite{0030804}, 
and it yields a fixed-parameter model checking algorithm for \msoone-definable properties on graphs of bounded cliquewidth~\cite{CourcelleMR00}.

It was conjectured by Courcelle~\cite{Courcelle90} that \msotwo-definability and recognizability for tree decompositions (i.e.~HR-recognizability) are equivalent for every graph class of bounded treewidth, 
provided that \msotwo is extended by counting predicates of the form ``the size of $X$ is divisible by $p$'',
for every integer $p$ (this logic is called \cmsotwo). This conjecture has been resolved by two of the current authors~\cite{bojanczyk2016definability}.
More precisely, in~\cite{bojanczyk2016definability} it was shown that for every $k$ there exists an \mso transduction which inputs a graph of treewidth at most $k$ and (nondeterministically) 
outputs its tree decomposition of
width bounded by a function of $k$. The graph is given via its incidence encoding. The conjecture of Courcelle then follows by composing this transduction with guessing the run of an automaton recognizing
the property in question
on the output decomposition.

The same question can be asked about cliquewidth: is it true that every class of graphs of bounded cliquewidth is \msoone-definable if and only if it is (VR-)recognizable? The present paper discusses this question, proving a special case of the equivalence.

\subsection*{Our contribution.} Our main result (Theorem~\ref{thm:main}) is that for every $k\in \Nats$,
there exists an \mso-transduction which inputs  a graph of \emph{linear} cliquewidth at most $k$, and outputs  a  cliquewidth decomposition of it which has width bounded by a function of $k$.
Here, we use the adjacency encoding of the graph. The {\em{linear cliquewidth}} of a graph is a linearized variant of cliquewidth, similarly as pathwidth is a linearized variant of treewidth;
see Section~\ref{sec:transductions} for definition and, e.g.,~\cite{adlkan15,GurskiW05,hegmeipap11,hegmeipap12} for more background. 
An immediate consequence of this result (Theorem~\ref{thm:recognizable}) is that every class of graphs 
of bounded linear cliquewidth is \cmsoone-definable if and only if it is \mbox{(VR-)recognizable}. This gives a partial answer to the question above.

The proof of our main result shares one key idea with the proof of the \msotwo-definability of tree decompositions, or more precisely, the pathwidth part of that proof \cite[Lemma 2.5]{bojanczyk2016definability}.
This is the use of Simon's Factorization Forest Theorem~\cite{Simon90}. We view a linear cliquewidth decomposition of width $k$ as a word over a finite alphabet and use the factorization theorem to construct a nested factorization of this word of depth bounded in terms of $k$.
The overall \mso transduction computing a decomposition is then constructed by induction on the nesting depth of this factorization. The technical challenge in this paper is to analyze the composition of ``subdecompositions'', which  is significantly more complicated in the cliquewidth case than in the treewidth/pathwidth case of \cite{bojanczyk2016definability}. In a path decomposition, each node of the path (over which we decompose) naturally corresponds to a separation of the graph, with the bag at the node being the separator. Thus, in the pathwidth case, each separation appearing in the decomposition essentially can be described by a tuple of vertices in the separator, with the left and the right side being essentially independent; this is a simple and easy to handle object. The difficulty in the cliquewidth case is that ``separations'' appearing in a linear cliquewidth decomposition are partitions of the vertex set into two sides with small ``modular complexity'': each side can be partitioned further into a bounded number of parts so that vertices from the same part have exactly the same neighbors on the other side. Such separations are much harder to control combinatorially, and hence capturing them using the resources of \mso requires a deep insight into the combinatorics of linear cliquewidth.

\section{Preliminaries}
\label{sec:transductions}

\subsection*{Graphs and cliquewidth.} 
All graphs considered in this paper are finite and simple. 
For the most part we use undirected graphs, if we use directed graphs then we remark this explicitly.
We write $[k]$ for $\{1,2,\ldots,k\}$ and $\binom{X}{1,2}$ for the family of nonempty subsets of a set $X$ of size at most 2.

A {\em{$k$-colored graph}} is  a graph with each vertex assigned a color from $[k]$.
On   $k$-colored graphs we define the following operations.
\begin{itemize}
\item {\bf Recolor.} For every function $\phi \colon [k]\to [k]$ there is a unary operation which inputs one $k$-colored graph and outputs the same graph where each vertex is recolored to the image
of its original color under $\phi$.
\item {\bf Join.} For every family of subsets $S\subseteq \binom{[k]}{1,2}$ there is an operation that inputs a family of $k$-colored graphs, of arbitrary finite size, and outputs a single $k$-colored graph constructed as follows.
Take the disjoint union of the input graphs and for each $\{i,j\}\in S$ (possibly $i=j$), add an edge between every pair of vertices that have colors $i$ and $j$, respectively, and originate from different 
input graphs.
\item {\bf Constant.} For each color $i\in [k]$ there is a constant which represents a graph on a single vertex with color $i$.
\end{itemize}
Define a \emph{width-$k$ cliquewidth decomposition} to be a (rooted) tree where nodes are labelled by operation names in an arity preserving way, that is, all constants are leaves and all recolor operations have exactly one child. The tree does not have any order on siblings, because Join is a commutative operation. For a cliquewidth decomposition, we define its \emph{result} to be the $k$-colored graph obtained by evaluating the operations in the decomposition. The {\em{cliquewidth}} of a graph is defined to be the minimum number $k$ for which there is a width-$k$ cliquewidth decomposition whose result is (some coloring of) the graph.

We remark that we somewhat diverge from the original definition of cliquewidth~\cite{CourcelleO00} in the following way. In~\cite{CourcelleO00}, there is one binary disjoint union operation that just adds two input $k$-colored graphs, and for each pair of different colors $i,j$ there is a unary 
operation that creates an edge between every pair of vertices of colors $i$ and $j$, respectively.
For our purposes, we need to have a union operation that takes an arbitrary number of input graphs.
Intuitively, this is because an \mso transduction constructing a cliquewidth decomposition cannot break symmetries and join isomorphic parts of the graph in some arbitrarily chosen order, which would be
necessary if we used binary disjoint union operations. More precisely, no \mso transduction can construct linear orders on vertices of edgeless graphs of unbounded size, while such an order can be \mso transduced from a cliquewidth decompositions with bounded-arity union operations. 

Another difference is that our join does simultaneously two operations: it takes the disjoint union of several inputs, and adds edges between them. When using binary joins, the two operations can be separated by introducing temporary colors; however when the number of arguments is unbounded such a separation is not possible. 

We note that the second feature described above (simultaneous disjoint union and addition of edges) also appears in the definition {\em{NLC-width}}~\cite{wanke1994k}, which is a graph parameter equivalent to cliquewidth, in the sense that every graph of NLC-width $k$ has cliquewidth at least $k$ and at most~$2k$~\cite{johansson1998clique}. Similarly to the proof of this fact, it can be easily shown that our definition of cliquewidth is at multiplicative factor at most $2$ from the original definition.

Linear cliquewidth is a linearized variant of cliquewidth, where we
allow only restricted joins that add only a single vertex.  More
precisely, we replace the Join and Constant operations with one unary
operation {\bf{Add Vertex}}. This operation is parameterized by a
color $i\in [k]$ and a color subset $X\subseteq [k]$, and it adds to
the graph a new vertex of color $i$, adjacent exactly to vertices with
colors belonging to $X$.  A \emph{width-$k$ linear cliquewidth
  decomposition} is a word consisting of Add Vertex and Recolor
operations, and the {\em{result}} of such a decomposition is the
$k$-colored graph obtained by starting with an empty graph, and applying the  the consecutive operations in the word from left to right. The {\em{linear cliquewidth}} of a graph is defined just like
cliquewidth, but we consider only linear cliquewidth decompositions. 
Note that we can transform any linear cliquewidth decomposition of width
$k$ to a cliquewidth decomposition of width at most $(k+1)$ by 
replacing each subterm 
$\textbf{Add Vertex}_{i,X}(\theta)$ by the term 
\[
  \textbf{Recolor}_{j\mapsto i}\Big(\textbf{Join}_{\{\{j,x\}\colon x\in X\}}\big(\theta,\textbf{Color}_j\big)\Big),
\]
where $j$ is a color not occurring in $\theta$. Hence the linear
cliquewidth of a graph is at least its cliquewidth minus
one.

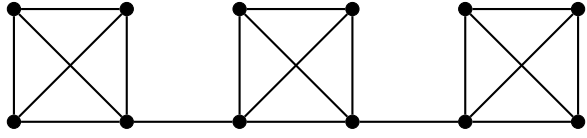
\begin{figure}
  \centering
  \begin{tikzpicture}[
    every node/.style={draw,fill,circle,minimum size = 5pt,inner sep=0pt},
    ]
    \foreach \x in {0,...,5}
    {
      \node (v\x) at (1.5*\x,0) {}; 
      \node (w\x) at (1.5*\x,1.5) {};
      \draw[thick] (v\x) edge (w\x); 
    }

    \foreach \x/\xx in {0/1,2/3,4/5}
    {
      \draw[thick] (v\x) edge (v\xx); 
      \draw[thick] (v\x) edge (w\xx); 
      \draw[thick] (w\x) edge (v\xx); 
      \draw[thick] (w\x) edge (w\xx); 
    }
    \draw[thick] (v1) edge (v2); 
    \draw[thick] (v3) edge (v4); 
  \end{tikzpicture}
  \caption{A graph of linear cliquewidth $3$}
  \label{fig:ex1}
\end{figure}

\begin{exa}
  Consider the graph $G$ displayed in Figure~\ref{fig:ex1}. 
  We will first argue that its cliquewidth is at most $3$, and then show that also its linear cliquewidth is at most $3$.

  We first construct a cliquewidth decomposition of $G$ of width 3. The 
  following three terms $\theta_1,\theta_2,\theta_3$ construct the
  three $4$-cliques of $G$ with appropriate colors:
  \begin{align*}
    \theta_1&=\textbf{Join}_{\{\{1\},\{1,2\}\}}\Big(\big\{\textbf{Color}_1, \textbf{Color}_1, \textbf{Color}_1, \textbf{Color}_2\big\}\Big),\\
\theta_2&=\textbf{Join}_{\{\{1\},\{1,2\},\{1,3\},\{2,3\}\}}\Big(\big\{\textbf{Color}_1,
          \textbf{Color}_1, \textbf{Color}_2,
          \textbf{Color}_3\big\}\Big),\\
    \theta_3&=\textbf{Join}_{\{\{1\},\{1,3\}\}}\Big(\big\{\textbf{Color}_1, \textbf{Color}_1, \textbf{Color}_1, \textbf{Color}_3\big\}\Big).
  \end{align*}
  Then the term
  \[
  \textbf{Join}_{\{\{2\},\{3\}\}}\left(\{\theta_1,\theta_2,\theta_3\}\right)
  \]
  is a width-$3$ cliquewidth decomposition of the graph $G$.

  For a linear cliquewidth decomposition, it is convenient to denote the unary operation
  $\textbf{Add Vertex}_{i,X}$ of
  adding a vertex of color $i$ and connecting it to all vertices of
  color $X$ by ${\boldsymbol a}_{i,X}$ and the recoloring operation by
  ${\boldsymbol r}_{\phi}$. This way, we may view a linear
  cliquewidth decomposition of width $k$ as a word over the finite
  alphabet
  \[
    \{{\boldsymbol a}_{i,X}\mid
    i\in[k],X\subseteq2^{[k]}\}\cup\{{\boldsymbol
      r}_{\phi}\mid\phi\colon [k]\to[k]\}.
  \]
  When given a word consisting of operations as above, we execute them in the order from left to right. Note that thus, the order of executing the operations in a word is {\em{opposite}} to the usual order in which function composition is denoted.
  With this notation, our linear cliquewidth decomposition of the graph $G$
  looks as follows:
  \[
\underbrace{{\boldsymbol a}_{1,\emptyset}{\boldsymbol a}_{1,\{1\}}{\boldsymbol a}_{1,\{1\}}{\boldsymbol a}_{2,\{1\}}}_{\text{first
    clique}}
\underbrace{{\boldsymbol a}_{3,\{2\}}{\boldsymbol r}_{\{2\mapsto1,3\mapsto2\}}{\boldsymbol a}_{2,\{2\}}{\boldsymbol a}_{2,\{2\}}{\boldsymbol a}_{3,\{2\}}{\boldsymbol r}_{\{2\mapsto1,3\mapsto2\}}}_{\text{second
    clique}}
\underbrace{{\boldsymbol a}_{3,\{2\}}{\boldsymbol a}_{3,\{3\}}{\boldsymbol a}_{3,\{3\}}{\boldsymbol a}_{3,\{3\}}}_{\text{third
    clique}}.
  \]
Again, recall here that operations in a linear cliquewidth decomposition are composed from left to~right.
\end{exa}

\subsection*{Relational structures and logic.}
Define a \emph{vocabulary} to be a set of \emph{relation names}, each one with associated arity in $\Nat$. 
A \emph{relational structure} over the vocabulary $\Sigma$ consists of a set called the \emph{universe}, and for each relation name in the vocabulary, an associated relation of the same arity over the universe. 
Note the possibility of relations of arity zero, such a relation stores a single bit of information about the structure.
A graph is encoded as a relational structure as follows: the universe is the vertex set, and there is one symmetric binary relation that encodes adjacency.

A width-$k$ cliquewidth decomposition of a graph is modeled as a relational structure 
whose universe is the set of nodes of the decomposition, there is a binary predicate ``child'', and for each operation from the definition of a cliquewidth decomposition there is a unary predicate (the set of these predicates depends on $k$) which selects nodes that use this operation. Note that the graph itself is not included in this structure, but it is straightforward to reconstruct it using an \mso transduction (see below).

To describe properties of relational structures, we use monadic second-order logic (\mso). 
This logic allows quantification both over single elements of the universe and also over subsets of the universe. 
For a precise definition of \mso, see~\cite{0030804}.
We will also use counting \mso, denoted also by \cmso, which is the extension of \mso with predicates of the form ``the size of $X$ is divisible by $p$'' for every $p \in \Nat$.

\subsection*{MSO transductions.}
We use the same notion of \mso transductions as in~\cite{bojanczyk2016definability,BojanczykP17}.
For the sake of completeness, we now recall the definition of an \mso transduction, which is taken verbatim from~\cite{BojanczykP17}.
We note that our \mso transductions differ syntactically from those
used in the literature, see e.g. Courcelle and
Engelfriet~\cite{0030804}, but are essentially the
same.

Suppose that $\Sigma$ and $\Gamma$ are finite vocabularies. 
Define a \emph{transduction} with input vocabulary~$\Sigma$ and output vocabulary $\Gamma$ to be a set of pairs 
\[\textrm{(input structure over $\Sigma$, output structure over $\Gamma$)}\]
which is invariant under isomorphism of relational structures.
Note that a transduction is a relation and not necessarily a function,
thus it can have many different possible outputs for the same input. 

An \mso transduction is any transduction that can be obtained by composing a finite number of {\em{atomic transductions}} of the following kinds.
Note that kind 1 is a partial function, kinds 2, 3, 4 are functions, and kind 5 is a relation.
\begin{enumerate}
	\item {\bf Filtering.} For every \mso sentence $\varphi$ over the input vocabulary there is transduction that discards structures where $\varphi$ is not satisfied and keeps the other ones intact. Formally, the transduction is the partial identity whose domain consists of the structures that satisfy the sentence. The input and output vocabularies are the same.
	\item {\bf Universe restriction.} For every \mso formula $\varphi(x)$ over the input vocabulary with one free first-order variable there is a transduction, which restricts the universe to those elements that satisfy $\varphi$. The input and output vocabularies are the same, the interpretation of each relation in the output structure is defined as the restriction of its interpretation 
	in the input structure to tuples of elements that remain in the universe.
	\item {\bfseries{\scshape{Mso}} interpretation.} This kind of transduction changes the vocabulary of the structure while keeping the universe intact. For every relation name $R$ of the output vocabulary, there is an \mso formula $\varphi_R(x_1,\ldots,x_k)$ over the input vocabulary which has as many free first-order variables as the arity of $R$. The output structure is obtained from the input structure by keeping the same universe, and interpreting each relation $R$ of the output vocabulary as the set of those tuples $(x_1,\ldots,x_k)$ that satisfy $\varphi_R$.
	\item {\bf Copying.} For  $k \in \set{1,2,\ldots}$, define $k$-copying to be the transduction which inputs a structure and outputs a structure consisting of $k$ disjoint copies of the input.
	Precisely, the output universe consists of $k$ copies of the input universe.
	The output vocabulary is the input vocabulary enriched with a binary predicate $\mathsf{copy}$ that selects copies of the same element, and unary predicates $\mathsf{layer}_1,\mathsf{layer}_2,\ldots,\mathsf{layer}_k$ which select elements belonging to the first, second, etc. copies of the universe.
	In the output structure, a relation name $R$ of the input vocabulary is interpreted as the set of all those tuples over the output structure, where the original elements of the copies were in relation $R$
	in the input structure.
	\item {\bf Coloring.} We add a new unary predicate to the input structure. Precisely, the universe as well as the interpretations of all relation names of the input vocabulary stay intact,
	but the output vocabulary has one more unary predicate. For every possible interpretation of this unary predicate, there is a different output with this interpretation implemented.
\end{enumerate}

\begin{exa}\label{ex:duplicate}
 Consider words over alphabet $\{a,b\}$ encoded as relational structures using two unary predicates $a$ and $b$ and one binary relation $\leq$ encoding the order in the word. The operation $w\mapsto ww$ of duplicating a word can be described as an \mso transduction from words to words as follows. First, we apply $2$-copying. Then, using an \mso interpretation, we define the order in the duplicated word as follows. For elements $x,y$, if $x$ and $y$ belong to the same copy of the universe, then they should be ordered as in the original word (i.e. we check the relation $\leq$ present in the vocabulary at this point). Otherwise, $x$ is before $y$ if and only if the predicates $\mathsf{layer}_1(x)$ and $\mathsf{layer}_2(y)$ hold. The interpretation leaves the predicates $a$ and $b$ intact. That is, for instance, predicate $a$ is interpreted using the formula $\varphi_a(x)=a(x)$.
 
 Note that in the \mso interpretation, we provide definitions only for the relations $a$, $b$, and $\leq$ present in the output vocabulary. Thus, the relations $\mathsf{copy}$, $\mathsf{layer}_1$, and $\mathsf{layer}_2$, which were introduced by the copying step, are effectively dropped in the interpretation step.
\end{exa}

Note that each element $v'$ of the output structure of an
\mso transduction is either identical to or a copy of an element $v$
of the input structure. We call this element $v$ the \emph{origin} of
$v'$. Thus we have a well-defined \emph{origin mapping} from the
output structure to the input structure. In general, this mapping is
neither injective nor surjective.

Define the \emph{size} of an atomic \mso transduction to be the size of its input and output vocabularies, plus the maximal quantifier rank of \mso formulas that appear in it (if the transduction uses \mso formulas). 
Define the {\em{size}} of an \mso transduction to be the sum of sizes of atomic transductions that compose to the transduction. 
Note that there are finitely many \mso transductions of a given size,
since there are finitely many \mso formulas (up to logical
equivalence) once the vocabulary, the free variables,
 and the quantifier rank are fixed.

An \mso transduction is \emph{deterministic} if it uses no
coloring. Note that a deterministic \mso transduction is a partial
function, that is, for each input structure there is at most one
output structure. For example, the transduction described in Example~\ref{ex:duplicate} is deterministic.

We remark that while our notion of an \mso transduction allows an arbitrarily long sequence of atomic transductions, of arbitrarily interleaving types, every \mso transduction can be reduced to the following normal form: first apply a finite sequence of coloring steps, then a single filtering step, then a single copying step, then a single \mso interpretation step, then a single universe restriction step, and finally, if necessary, apply an \mso interpretation that just renames some relations. For this, see Theorem~4 in~\cite{arxiv-BojanczykP17} or the discussion in~\cite{0030804}, where a similar kind of a normal form is assumed as the original definition of an \mso transduction. We will not use this fact in this work.

\medskip

Note that the composition of two \mso transductions is an \mso transduction by definition.
Another well-known property that we will use, as expressed in the following lemma, is that the union of two \mso transductions is also an \mso transductions; 
recall that here we regard \mso transductions as relations between input and output structures.
This property is Lemma~7.18 from~\cite{0030804}. 
Since our notion of an \mso transduction is a bit different from the one used in~\cite{0030804}, we give a proof for completeness.

\begin{lem}\label{lem:union-trans}
The union of two \mso transductions with the same input and output vocabularies is also an \mso transduction.
\end{lem}
\begin{proof}
First, using copying create two copies of the universe, called further the first and the second {\em{layer}}.
Apply the first transduction only to the first layer, thus turning it into a (nondeterministically chosen) result of the first transduction.
More precisely, all formulas used in the first transduction are relativized to the first layer, or any new elements originating from them. 
Also, each copying step is followed by an additional universe restriction step that removes unnecessary copies of the second layer.
Then, analogously apply the second transduction only to the second layer.
After this step, the structure is a disjoint union of some result of the first transduction applied to the initial structure, and some result of the second transduction applied to the initial structure.
It remains to nondeterministically choose one of these results using coloring, and remove the other one using universe restriction.
\end{proof}

The key property of \mso transductions is that \cmso- and \mso-definable properties are closed under taking inverse images over \mso transductions.
More precisely, we have the following statement.

\begin{lem}[Backwards Translation Theorem,~\cite{0030804}]
Let $\Sigma,\Gamma$ be finite vocabularies and let $\mathcal{I}$ be an \mso transduction with input vocabulary $\Sigma$ and output vocabulary $\Gamma$.
Then for every \mso (resp. \cmso) sentence $\psi$ over $\Gamma$ there exists an \mso (resp. \cmso) sentence $\varphi=\mathcal{I}^{-1}(\psi)$ over $\Sigma$ such that $\varphi$ holds 
in exactly those $\Sigma$-structures on which $\mathcal{I}$ produces at least one output satisfying $\psi$.
\end{lem}

Finally, let us point out that \mso transductions can be used to give another explanation of the connection between the \mso logic and graph parameters such as cliquewidth and treewidth, through interpretability in trees. Namely, it is known that a class of graphs has uniformly bounded cliquwidth if and only if it is contained in the image of the class of trees under a fixed \mso transduction. Similarly, a class of graphs has uniformly bounded treewidth if and only if their incidence graphs are contained in the image of the class of trees under a fixed \mso transduction. See~\cite{0030804} for a broader discussion.

\subsection*{Simon's Lemma.}
As we mentioned in Section~\ref{sec:intro}, the main technical tool used in this work will be Simon's Factorization Theorem~\cite{Simon90}.
We will use the following variant, which is an easy corollary of the
original statement. Recall that a \emph{semigroup} is an algebra with one
associative binary operation, usually denoted as multiplication, and
that an {\em{idempotent}} in a semigroup is an element $e$ such that
$e\cdot e=e$.

\begin{lem}[Simon's Lemma]\label{lem:Simon}
Suppose that $S$ and $T$ are semigroups, where $S$ is finitely generated (but possibly infinite) and $T$ is finite. 
Suppose further that $h\colon S\to T$ is a semigroup homomorphism and $f\colon \Nat\to \Nat$ and $\mu\colon S\to \Nat$ 
are functions such that 
\begin{equation}\label{eq:Simon}
  \mu(s_1 \cdot \ldots\cdot s_n) \leq f(\max_{i\in [n]}\, \mu(s_i))
\end{equation}
holds whenever $n=2$ or there is some idempotent $e \in T$ such that $e=h(s_1)=\ldots=h(s_n)$. Then $\mu$ has finite range, i.e.~there exists $K\in \Nat$ such that $\mu(s)\leq K$ for all $s\in S$.
\end{lem}

Before we proceed to the proof of Simon's Lemma,
we first recall the original statement of Simon's Factorization Theorem, as described by Kufleitner~\cite{Kufleitner08}.
Suppose $\Sigma$ is a finite alphabet and let $\Sigma^+$ be the semigroup of nonempty words over $\Sigma$ with concatenation.
Suppose further we are given a finite semigroup $T$ and a homomorphism $h\colon \Sigma^+\to T$.

For a word $u\in \Sigma^+$ of length more than $1$, we define two types of factorizations:
\begin{itemize}
\item {\bf{Binary}}: $u=u_1u_2$ for some $u_1,u_2\in \Sigma^+$, and
\item {\bf{Idempotent}}: $u=u_1\cdots u_n$ for some $u_1,\ldots,u_n\in \Sigma^+$ such that all words $u_i$ have the same image under $h$, which is moreover an idempotent in $T$.
\end{itemize}
Define the {\em{$h$-rank}} of a word $u\in \Sigma^+$ as follows.
If $u$ has length $1$ then its $h$-rank is $1$.
Otherwise, we define the $h$-rank of $u$ as
\[1+\min_{u=u_1\ldots u_n}\, \max_{i\in [n]}\ \textrm{$h$-rank of }u_i,\]
where the minimum is over all binary or idempotent factorizations of $u$. 
Simon's Factorization Theorem can be then stated as follows.

\begin{thm}[Simon's Factorization Theorem, \cite{Kufleitner08,Simon90}]\label{thm:Simon}
If $h$ is a homomorphism from $\Sigma^+$ to a finite semi-group $T$, then every word from $\Sigma^+$ has $h$-rank at most $3|T|$.
\end{thm}

The existence of an upper bound expressed only in terms of $|T|$ was first proved by Simon~\cite{Simon90}, while the improved upper bound of $3|T|$ is due to Kufleitner~\cite{Kufleitner08}. We proceed to the proof of our Simon's Lemma.

\begin{proof}[Proof of Simon's Lemma]
Let $\Sigma=\{g_1,\ldots,g_k\}$ be the generators of $S$, 
and let $\iota\colon \Sigma^+\to S$ be the natural homomorphism that computes the product of sequences of generators in $S$.
Consider the homomorphism $h'\colon \Sigma^+\to T$ defined as the composition of $\iota$ and $h$.

We prove the following claim: for each $d\in \Nats$ there is a number $K_d$ such that for every word $u\in \Sigma^+$ of $h'$-rank at most $d$, we have $\mu(\iota(u))\leq K_d$.
Observe that this will finish the proof for the following reason. 
By Theorem~\ref{thm:Simon}, every element $s\in S$ can be expressed as $s=\iota(u)$ for some $u\in \Sigma^+$ of $h'$-rank at most $3|T|$.
Hence we can take $K=K_{3|T|}$.

We prove the claim by induction on $d$. For $d=1$ we have that $u$ has to consist of one symbol, so we can take 
\[K_1=\max_{i\in [k]} \mu(h(g_i)).\]
Suppose then that $d>2$ and take any word $u$ of $h'$-rank equal to $d$. By the definition of the $h'$-rank, $u$ admits a factorization $u=u_1\ldots u_n$ into factors $u_i$ of $h'$-rank smaller than $d$,
such that either $n=2$, or all words $u_i$ have the same image under $h'$, which is moreover an idempotent in $T$.
By the supposition of the lemma and induction assumption, we have
\[\mu(\iota(u_1\cdots u_n))=\mu(\iota(u_1)\cdots\iota(u_n))\leq f(\max_{i\in [n]}\, \mu(\iota(u_i)))\leq f(K_{d-1}).\]
Hence we can take $K_d=\max(K_{d-1},f(K_{d-1}))$.
\end{proof}

\section{Main results}

\subsection*{Statement of the main result.}
Our main result is that for every $k$, there is an \mso transduction which maps every graph of linear cliquewidth $k$ to some of its cliquewidth decompositions. The width of these decompositions is bounded by a function of $k$; we do not achieve the optimal value $k$.
To state this result, we introduce a graph parameter, called \emph{definable cliquewidth}, which measures the size of an \mso transduction necessary to transform the graph into its cliquewidth decomposition. 


Recall that we model a width-$k$ cliquewidth decomposition of a graph as a (rooted) tree labelled by an alphabet of operations depending on $k$. Such a cliquewidth decomposition $t$ constructs a graph $G_t$ whose vertices are the leaves of the tree. More generally, we say that $t$ is a cliquewidth decomposition of a graph $G$ if there is an isomorphism from $G_t$ to $G$.

\begin{defi}[Decomposer]\label{def:decomposer}
A \emph{width-$k$ decomposer} is  an \mso transduction $\mathcal{D}$
from the vocabulary of graphs to the vocabulary of width-$k$ cliquewidth decompositions
such that for every input-output pair $(G,t)$ of $\mathcal D$ the following two conditions are satisfied.
\begin{enumerate}[(a)]
\item\label{c:correct} $t$ is a width-$k$ cliquewidth decomposition of $G$.
\item\label{c:origin} The origin mapping from $t$ to $G$ restricted to the leaves of $t$ is an isomorphism from $G_t$ to~$G$.
\end{enumerate}
\end{defi}

Condition \ref{c:origin} in the definition of decomposers may seem unnecessarily restrictive, but in fact will turn out to be very useful in the technical arguments (see Section~\ref{sec:toolbox}). Furthermore, natural transductions satisfying \ref{c:correct} also tend to satisfy \ref{c:origin}, because usually such transduction proceed by building the tree of a cliquewidth decomposition on top of the input graph.

Note that the \emph{size} of a decomposer (as a particular \mso-transduction) is an upper bound for its width, because the size of a transduction is larger than the size of its output vocabulary.

\begin{defi}[Definable cliquewidth] 
The \emph{definable cliquewidth} of a graph $G$, denoted by $\dcw(G)$, is the smallest size of a decomposer which produces at least one output on $G$.  
\end{defi}

Note that the decomposer witnessing the value of the definable cliquewidth of a graph $G$, when applied to $G$, outputs (at least one) cliquewidth decomposition of $G$ of width at most $\dcw(G)$. This is because the width of a decomposer is bounded by its size.

Observe also that there are finitely many decomposers of a given size, and decomposers are closed under union by Lemma~\ref{lem:union-trans}. 
Therefore, for every $k$ there is a single decomposer (of width $k$ and size $f(k)$) 
which produces at least one output on every graph with definable cliquewidth at most $k$, namely one can take the union of 
all decomposers of size at most $k$.

The main result of this paper is the following.

\begin{thm}\label{thm:main}
For every $k\in \Nat$ there exist a decomposer $\mathcal D$ that for every graph $G$ of linear cliquewidth at most $k$ produces at least one output. 
In particular, the definable cliquewidth of a graph is bounded by a function of its linear cliquewidth.  
\end{thm}

The result above could be improved in two ways: first, we could make the transduction produce results for graphs of bounded cliquewidth (and not bounded linear cliquewidth), and second, we could produce cliquewidth decompositions of optimum width. We leave both of these improvements to future work. Note that it is impossible to find a decomposer which produces a linear cliquewidth decomposition for every graph of linear cliquewidth $k$; the reason is that such a decomposer would impose a total order on the vertices of the input graph, and this is impossible for some graphs, such as large edgeless graphs.

We remark that, similarly to the case of treewidth~\cite{bojanczyk2016definability}, our proof is effective:
the decomposer $\mathcal D$ can be computed from $k$. This essentially follows from a careful inspection of the proofs, so we usually omit the details in order not to obfuscate the main ideas with computability issues
of secondary importance. There is, however, one step in the proof (Lemma~\ref{lem:induced-dcw}) where computability of a bound is non-trivial, hence there we present an explicit discussion.


\subsection*{Recognizability.} 
We now state an important corollary of the main theorem, namely that for graph classes with linear cliquewidth, being definable in (counting) \mso is the same as being recognizable. Let us first define the notion of recognizability that we use.
 For $k \in \Nat$, define a \emph{$k$-context}
to be a width-$k$ 
cliquewidth decomposition
with one distinguished leaf. If $t$ is a $k$-context and $G$ is a $k$-colored graph, then $t[G]$ is defined to be the $k$-colored graph obtained by replacing the distinguished leaf of $t$ by $G$, and then applying all the operations in $t$.
 
 \begin{defi}[Recognizability, see \cite{0030804}, Def.~4.29]\label{def:recog}
 Let  $L$ be a class of graphs. Two $k$-colored graphs $G_1,G_2$ are called $L$-equivalent if for every $k$-context $t$ we have $t[G_1] \in L$ iff $t[G_2] \in L$,
where membership in $L$ is tested after ignoring the coloring.
We say that $L$ is \emph{recognizable} if for every $k \in \Nat$ there are finitely many equivalence classes of $L$-equivalence.
 \end{defi}

Theorem 5.68(2) in~\cite{0030804} shows that if a class of graphs is definable in \mso (in the sense used here; this logic is also called \msoone in~\cite{0030804}),
then it is recognizable (in the sense of Definition~\ref{def:recog}, which is called~{\em{VR-recognizable}} in~\cite{0030804}). 
The converse implication is not true, e.g.,~there are uncountably many recognizable graph classes. 
The following result, which is a corollary of our main theorem, says that the converse implication is true under the assumption of bounded linear cliquewidth. 

\begin{thm}\label{thm:recognizable}
If $L$ is a class of graphs of bounded linear cliquewidth, then $L$ is recognizable if and only if it is definable in \cmso.
\end{thm}
\begin{proof}
As mentioned above, the right-to-left implication is true even without assuming a bound on linear cliquewidth. For the converse, we use the following claim; since the proof is completely standard, we only sketch it.

\begin{clm}\label{cl:rec-def}
If a class of graphs $L$ is recognizable, then for every $k$ the following language $L_k$ of labelled trees is definable in \cmso.
\begin{align*}
L_k=\{ t \colon \mbox{$t$  is a tree that is a width-$k$ cliquewidth decomposition}\\
\mbox{whose resulting graph is in $L$}\}
\end{align*}
\end{clm}
\begin{proof}[Proof sketch]
The language $L_k$ is a set of (unranked) trees without sibling order.
Define $\tilde{L}_k$ to be the language of sibling-ordered trees such that if the sibling order is ignored, then the resulting tree belongs $L_k$. 
Using the assumption that $L$ is recognizable, one shows that $\tilde{L}_k$ is definable in \mso; the idea is that using the sibling order an \mso formula can convert a tree into one which has binary branching, 
and then compute for each subtree its $L$-equivalence class.  
As shown in~\cite{Courcelle90}, if a language of sibling-ordered trees is definable in \mso and invariant under reordering siblings, then the language of sibling-unordered trees obtained from it by ignoring the
sibling order is definable in \cmso without using the sibling order.
Applying this to $\tilde{L}_k$ and $L_k$ we obtain the claim.
\cqed\end{proof}
Using Claim~\ref{cl:rec-def}, we complete the left-to-right implication. Assume every graph from $L$ has linear cliquewidth at most $k$. Apply Theorem~\ref{thm:main}, yielding a decomposer $\mathcal{D}$ from graphs to width-$\ell$ cliquewidth decompositions which produces at least one output for every graph in~$L$. Apply Claim~\ref{cl:rec-def} to $L$ and~$\ell$. 
Since $\mathcal{D}$ produces at least one output for every graph in~$L$, we have that $L$ is the inverse image under $\mathcal{D}$ of the language $L_\ell$ in the conclusion of the claim. 
It follows from the Backwards Translation Theorem that $L$ is definable in \cmso.
\end{proof}

\newcommand{\Cells}{\mathcal{C}_k}

\section{The proof strategy}
\label{sec:simon-strategy-der}
In this section we present the proof strategy for our main contribution, Theorem~\ref{thm:main}. 

A linear cliquewidth decomposition of width $k$, being a single path, can be viewed as  a sequence of instructions. For such sequences of instructions (actually, for a similar but slightly more general object), we will use the name \emph{$k$-derivations}.  Intuitively speaking, a $k$-derivation corresponds to an infix of a linear cliquewidth decomposition of width $k$, and it represents the total history of operations performed in this infix. This history consists of the constructed (colored) subgraph constructed and the composition of applied recolorings. 
We can concatenate $k$-derivations, which means that the set of $k$-derivations is endowed with a semigroup structure. 
The main idea is to use Simon's Factorization Theorem~\cite{Simon90}, in the flavor delivered by the Simon's Lemma (Lemma~\ref{lem:Simon}), to factorize this product into a tree of bounded depth, so that definable cliquewidth decompositions of factors can be constructed via a bottom-up induction over the factorization.

More precisely, the Simon's Lemma is used to prove Theorem~\ref{thm:main} as follows. As the semigroup $S$ we use $k$-derivations. 
As the homomorphism $h$, we use a notion of {\em{abstraction}}, which maps each  $k$-derivation to a bounded-size combinatorial object consisting of all the information we need to remember about it.
Composing $k$-derivations naturally corresponds to composing their abstractions, which formally means that the set of abstractions, whose size is bounded in terms of $k$,
can be endowed with a semigroup structure so that taking an abstraction of a $k$-derivation is a semigroup homomorphism.
By taking $\mu$ to be the definable cliquewidth of a graph, we use the Simon's Lemma to show that $\mu$ has a finite range on the set of all $k$-derivations, i.e.~there is a finite upper bound on the definable cliquewidth of all $k$-derivations.
To this end, we need to prove that the assumptions of the Simon's Lemma are satisfied, that is, condition~\eqref{eq:Simon} is satisfied when either $n=2$ or  all the abstractions of all $k$-derivations
in the product are equal to some idempotent in the semigroup of abstractions.

We now set off to implement this plan formally. For the rest of the paper we fix $k\in \Nat$. Our goal is to show that graphs of linear cliquewidth at most $k$ have bounded definable cliquewidth.

\subsection*{Derivations.} 
\begin{figure}[t]
	\begin{center}
		\includegraphics[page=1,scale=0.65]{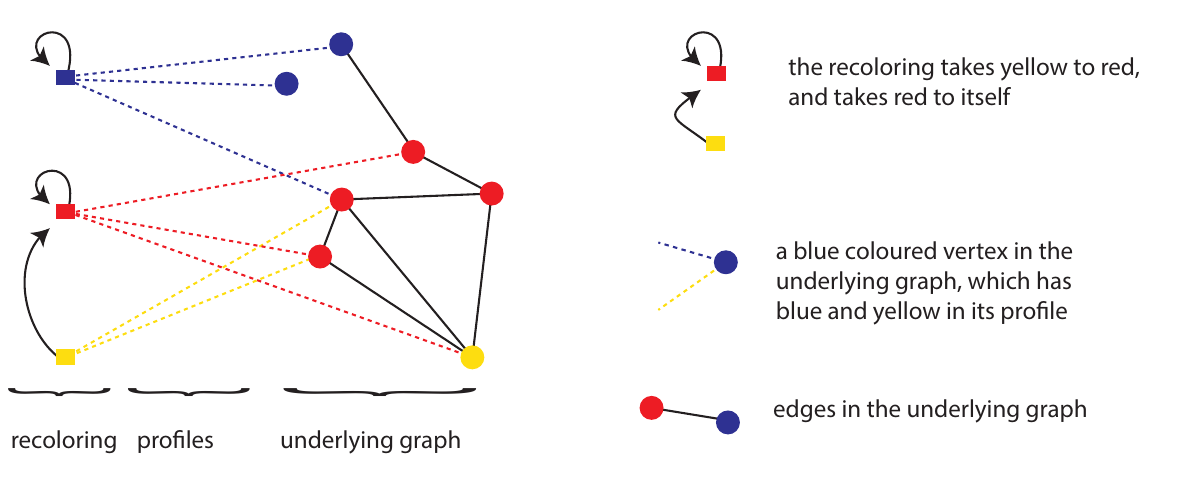}
		\caption{A $k$-derivation for $k=3$, with the numbers $\set{1,2,3}$ being represented as colors $\set{\text{red, blue, yellow}}$.  The boxes indicate colors as used by the profiles and recoloring, and the circles indicate vertices of the underlying graph.}\label{fig:dupa}
	\end{center}
\end{figure}

We first introduce $k$-derivations and their semigroup.
\begin{defi}
A $k$-derivation $\sigma$ is a triple $(G,\lambda,\phi)$, where
\begin{itemize}
\item $G$ is a $k$-colored graph, called the {\em{underlying graph of $\sigma$}};
\item $\lambda\colon V(G)\to 2^{[k]}$ is a function that assigns to each vertex $u$ its {\em{profile}} $\lambda(u)\subseteq [k]$; and
\item $\phi\colon [k]\to [k]$ is a function called the {\em{recoloring}}.
\end{itemize}
\end{defi}
Intuitively, if we treat a $k$-derivation $\sigma=(G,\lambda,\phi)$ 
as a subword of instructions in a linear cliquewidth decomposition, then $G$ is the subgraph induced by vertices introduced by these instructions and $\phi$ is the composition of all recolorings applied.
The profile $\lambda$ has the following meaning: supposing there were some instructions preceding the $k$-derivation in question, it assigns each vertex $u$ of $G$ a subset $\lambda(u)$ of colors such 
that among vertices introduced by these preceding instructions,
$u$ is adjacent exactly to vertices with colors from~$\lambda(u)$. See Figure~\ref{fig:dupa}
for an example.

By the definable cliquewidth of a $k$-derivation we mean the definable cliquewidth of its underlying graph, with the colors ignored.
For a $k$-derivation $\sigma=(G,\lambda,\phi)$ and  $c=(i,X)\in [k]\times 2^{[k]}$, the set of all vertices with color $i$ and profile $X$ is be called the {\em{$c$-cell}}, and denoted by $\sigma[c]$. 
For brevity, we write $\Cells=[k]\times 2^{[k]}$ and interpret $\Cells$ as the index set of cells in $k$-derivations.
By  abuse of  notation, we use the term {\em{cell}} also for the elements of $\Cells$.

\begin{figure}[t]
	\begin{center}
		\includegraphics[page=2,scale=0.65]{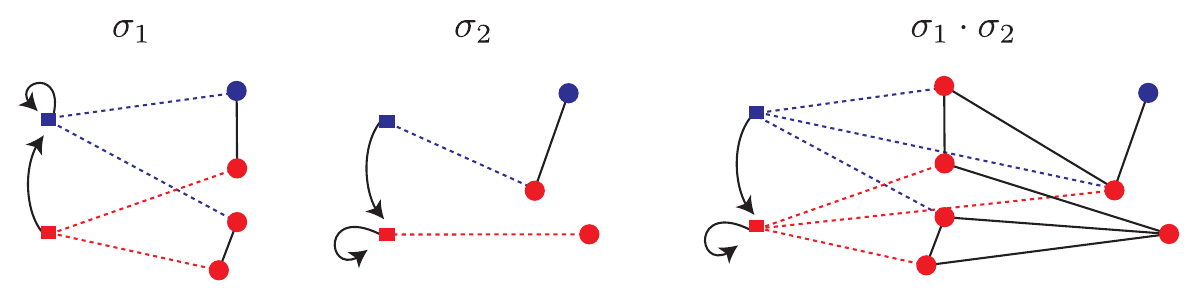}
		\caption{Example composition of derivations, depicted using the convention from Figure~\ref{fig:dupa}. Note that in the composed derivation $\sigma_1\cdot \sigma_2$, the recoloring is the composition of recolorings from $\sigma_1$ and $\sigma_2$, while the underlying graph is obtained from the disjoint union of the underlying graphs of $\sigma_1$ and $\sigma_2$ by appropriately adding edges according to colors in $\sigma_1$ and profiles in $\sigma_2$. Colors and profiles of vertices are essentially inherited from $\sigma_1$ and $\sigma_2$, except that the colors of vertices featured in $\sigma_1$ are updated using the recoloring from $\sigma_2$, while profiles of vertices featured in $\sigma_2$ are modified to take into account the recoloring from $\sigma_1$.}\label{fig:compo}
	\end{center}
\end{figure}

We now describe the semigroup structure of $k$-derivations.
We define the {\em{composition}} $\sigma_1\cdot \sigma_2$ of  two $k$-derivations $\sigma_1=(G_1,\lambda_1,\phi_1)$ and $\sigma_2=(G_2,\lambda_2,\phi_2)$ as follows; see Figure~\ref{fig:compo} for an illustration.
The underlying graph of the composition is constructed by taking the disjoint union of $\phi_2(G_1)$ and $G_2$, where $\phi_2(G_1)$ denotes $G_1$ with the color of each vertex substituted with its image under $\phi_2$,
and adding an edge between a vertex $u\in G_1$ and a vertex $v\in G_2$ whenever the color of $u$ in $G_1$ belongs to the profile $\lambda_2(v)$.
The profile of a vertex $u$ in the composition is equal to $\lambda_1(u)$ if $u$ originates from $G_1$, and to $\phi_1^{-1}(\lambda_2(u))$ if $u$ originates from $G_2$.
Finally, the recoloring in the composition is the composition of recolorings, that is, $\phi_2\circ \phi_1$.
It is straightforward to see that composition is associative, and hence it turns the set of $k$-derivations into a semigroup. Let us stress here that derivations are composed from left to right, similarly to operations in linear cliquewidth decompositions.

Define an \emph{atomic} $k$-derivation to be one where the underlying graph has at most one vertex. 
The number of different atomic $k$-derivations is finite and bounded only in terms of~$k$, because the only freedom is the choice of the color and the profile of the unique vertex (if there is one), 
as well as the recoloring.
Define $S_k$ to be the subsemigroup of the semigroup of all $k$-derivations which is  generated by the atomic $k$-derivations. By definition, $S_k$ is finitely generated.
The next lemma is a straightforward reformulation of the definition of linear cliquewidth.

\begin{lem}\label{lem:redefine-clique-width}
If a graph has linear cliquewidth at most $k$, then it is the underlying graph of some $k$-derivation $\sigma \in S_k$.
\end{lem}
\begin{proof}
Take any width-$k$ linear cliquewidth decomposition of the graph $G$ in question and turn it into a sequence of atomic $k$-derivations as follows. 
Every $\mathbf{Recolor}_{\phi}$ operation is replaced with an atomic $k$-derivation 
with empty underlying graph and recoloring $\phi$, whereas every $\mathbf{Add Vertex}_{i,X}$ operation is replaced by an atomic $k$-derivation with identity recoloring and
underlying graph consisting of one vertex of color $i$ and profile $X$. The composition of the obtained sequence
yields a $k$-derivation $\sigma\in S_k$ whose underlying graph is $G$.
\end{proof}

\subsection*{Abstractions.} Our goal is to apply the Simon's Lemma to the finitely generated semigroup $S_k$, with $\mu$ being the definable cliquewidth of the underlying graph. 
To apply the Simon's Lemma, we also need a homomorphism from $S_k$ to some finite  semigroup. This homomorphism is going to be abstraction, and we define it below.

To define the abstraction of a derivation, we need one more auxiliary concept, namely the  {\em{flipping}} of a graph.
For a graph $G$ and vertex subsets $X,Y\subseteq V(G)$, the {\em{flip between $X$ and $Y$}} is defined to be the following operation modifying $G$: for each $x\in X$, $y\in Y$, $x\neq y$, if
there is an edge $xy$ then remove it, and otherwise add it. In other words, flipping between $X$ and $Y$ means reversing the adjacency relation in all pairs of different elements from $X\times Y$.
Note that in the flip operation, the sets $X$ and $Y$ need not be disjoint. 
Suppose  that $\sigma$ is a $k$-derivation. Recall that $\Cells$ represents the names of cells, i.e.~each element of $\Cells$ is a pair (vertex color, profile). For a subset $Z\subseteq \binom{\Cells}{1,2}$ define  the {\em{$Z$-flip of $\sigma$}} to be the graph obtained from the underlying graph of $G$ by performing  the flip between $\sigma[c]$ and $\sigma[d]$ for each $\{c,d\}\in Z$. Note that $Z$ can contain singletons, i.e.~we might have $c=d$.

\begin{defi}\label{def:derivation}
For a $k$-derivation $\sigma$, its {\em{abstraction}}, denoted by $\abst{\sigma}$, is the triple $(L,\rho,\phi)$ consisting of the following information about $\sigma$:
\begin{itemize}
\item $L\subseteq \Cells$ is the set of cells that are non-empty in $\sigma$, called {\em{essential}};
\item $\rho\subseteq 2^{\binom{\Cells}{1,2}}\times \Cells\times \Cells\times 2^{\Cells}$ is the {\em{connectivity registry}}, which contains all tuples $(Z,c,d,W)$ such that: in the $Z$-flip of $\sigma$ there is a path that starts in a vertex of $\sigma[c]$, ends in a vertex of $\sigma[d]$, and all of whose
 internal vertices belong to $\bigcup_{b\in W} \sigma[b]$;
\item $\phi$ is the recoloring function of $\sigma$.
\end{itemize}
\end{defi}

We briefly explain the idea behind the connectivity registry. In general, we would like to remember which pairs of cells can be connected by a path in the underlying graph of the derivation. 
However, throughout the proof, particularly in Section~\ref{app:def-ord}, we will often working not with a $k$-derivation, but with some $Z$-flip of it. Therefore, we want the abstraction to store the connectivity information after every possible flip. For technical reasons, we also remember the subset of cells that are traversed by the path.

Denote by $T_k$ the set of all possible abstractions of $k$-derivations; note that $T_k$ is a finite set whose size depends only on $k$, albeit it is doubly exponential in $k$.
We leave it to the reader to prove that ``having the same abstraction'' is a congruence in the semigroup $S_k$, that is, an equivalence relation $\sim$ on $S_k$ such that $s\sim s'$ and $t\sim t'$ imply $st\sim s't'$ for all $s,s',t,t'\in S_k$.
 It follows that we may endow $T_k$ with a unique binary composition operation which makes it into a semigroup, and which makes the abstraction function a semigroup homomorphism from $S_k$ to $T_k$.

\subsection*{Applying the Simon's Lemma.} We will apply the Simon's Lemma for $S=S_k$, $T=T_k$, $h$ being the abstraction operation, and $\mu$ being the definable cliquewidth of 
the underlying  graph of a $k$-derivation (after forgetting the coloring). The conclusion of the Simon's Lemma will say that $\mu$ has bounded range, i.e.~there is a finite bound on the definable cliquewidth of  the underlying graphs of derivations from $S_k$. Since these underlying graphs are the same as graphs of linear cliquewidth at most $k$ by Lemma~\ref{lem:redefine-clique-width}, this will mean that bounded linear cliquewidth implies bounded definable cliquewidth, thus proving Theorem~\ref{thm:main}.

To apply the Simon's Lemma,  we need to verify that assumption~\eqref{eq:Simon} is satisfied for some function $f\colon \Nat\to \Nat$.
The treatment of cases when $n=2$, and when all derivations have a common idempotent abstraction, is different, as encapsulated in the following two lemmas.

\begin{lem}[Binary Lemma]
There is a function $f\colon \Nat\to \Nat$ such that \[\dcw(\sigma\cdot \tau)\leq f(\max(\dcw(\sigma),\dcw(\tau)))\] for every $\sigma,\tau\in S_k$.
\end{lem}

\begin{lem}[Idempotent Lemma]
There is a function $f\colon \Nat\to \Nat$ such that 
\[\dcw(\sigma_1\cdots \sigma_n)\leq f(\max_{i\in [n]}\, \dcw(\sigma_i))\]
for every $\sigma_1,\ldots,\sigma_n$ which have the same abstraction, and this abstraction is idempotent. 
\end{lem}

Condition~\eqref{eq:Simon} of the Simon's Lemma then follows by taking $f$ to be the maximum of the functions given by the Binary and the Idempotent Lemma.
Thus, we are left with proving these two results.
The proof of the Binary Lemma is actually quite easy and we could present it right away, but it will be more convenient to use technical tools developed in the proof of the Idempotent Lemma, so we postpone it to Section~\ref{sec:toolbox}.

\newcommand{\bkl}{\prec}
\newcommand{\bkg}{\succ}
\newcommand{\bkleq}{\preceq}
\newcommand{\bkgeq}{\succeq}
\newcommand{\bkeq}{\equiv}
\newcommand{\Ga}{\widehat{G}}
\newcommand{\Ha}{\widehat{H}}

\newcommand{\Af}{\mathfrak{A}}
\newcommand{\Bf}{\mathfrak{B}}

\section{Proof of the Idempotent Lemma}\label{sec:idempotent}

In this section we prove the Idempotent Lemma assuming a technical result called the Definable Order Lemma, which we will explain in a moment. 
Let us consider a sequence $\sigma_1,\ldots,\sigma_n$ of $k$-derivations such that for some abstraction $e$ that is idempotent in $T_k$,
we have $e=\abst{\sigma_1}=\ldots=\abst{\sigma_n}$.
Let $\sigma=\sigma_1\cdots \sigma_n$, and let $G$ be the underlying graph of $\sigma$. Moreover, for $i\in [n]$ the underlying graph of $\sigma_i$ shall be denoted by $G_i$, and we call it also the {\em{$i$-th block}}.

Let $\bkleq$ be the linear quasi-order (i.e.~a total, transitive and reflexive relation) defined on the vertex set of $G$ as follows: 
$u \preceq v$ holds if and only if $u$ belongs to the $i$-th block and $v$ belongs to the $j$-th block for some $i \le j$.
Similarly, let $\bkeq$ be the equivalence relation on the vertex set of $G$ defined as belonging to the same block; that is, $u\bkeq v$ iff $u\bkleq v$ and $v\bkleq u$.
The relations $\bkleq$ and $\bkeq$ will be called the {\em{block order}} and the {\em{block equivalence}}, respectively.
Our general idea is to show that the block order, and hence also the block equivalence, can be interpreted using a bounded size (nondeterministic) \mso formula, i.e.~that it has bounded (in terms of $k$) interpretation complexity as defined below.

\begin{defi}[Interpretation complexity] 
Suppose that $\structa$ is a relational structure, and let $R$ be a relation on its universe, say of arity $n$. 
Define the \emph{interpretation complexity} of $R$ inside $\structa$ to be the smallest $m$ 
such that there exist subsets $X_1,\ldots,X_m$  of the universe in $\structa$ and an \mso formula $\varphi(x_1,\ldots,x_n,X_1,\ldots,X_m)$ of quantifier rank at most $m$
over the vocabulary of $\structa$ such that
\begin{align*}
  (x_1,\ldots,x_n) \in R \quad \mbox{iff}\quad \varphi(x_1,\ldots,x_n,X_1,\ldots,X_m) \qquad\mbox{for all $x_1,\ldots,x_n$ in $\structa$.}
\end{align*}
\end{defi}

If the interpretation complexity of the block order was bounded by a function of $k$, then we would construct a cliquewidth decomposition of $G$ as follows: 
first construct cliquewidth decompositions of all blocks, and then combine them sequentially along the block order.
Unfortunately, in general we cannot hope for such a bound.  
To see this, consider the example where $G$ consists of, say, two disjoint paths of length $n$ each, plus an independent set of size $n$. 
In this example, each $\sigma_i$ introduces the $i$-th vertex of each of the two paths and one vertex in the independent set. 
It is not difficult to see that in this example the interpretation complexity of the block order grows with the number of blocks. However, we can define the block order on each connected component (i.e.~each of the two paths, and each vertex of the independent set) separately, and a cliquewidth decomposition of the whole graph can be obtained by putting a Join over decompositions of components.
Thus, the obtained decomposition will have a different shape than the input linear decomposition corresponding to the product $\sigma_1\cdots \sigma_n$.
The following statement, which is our main technical result towards the proof of the Idempotent Lemma, explains how this plan can be implemented in general.

\begin{lem}[Definable Order Lemma]\label{lem:definable-order}
Let $\sigma_1,\ldots,\sigma_n$ be $k$-derivations as in the assumption of the Idempotent Lemma. 
There exists a set $Z\subseteq \binom{\Cells}{1,2}$ such that if $\sim$ is the relation of being in the same connected component in the $Z$-flip of $\sigma_1 \cdots \sigma_n$, 
then the relation  $\sim \cap \preceq$ has interpretation complexity over $G$ bounded by a function of $k$.
\end{lem}

For now we postpone the proof of the Definable Order Lemma; it will be presented in Section~\ref{app:def-ord}.
In the rest of this section we show how to use this result to prove the Idempotent Lemma.
Along the way we will develop a relevant toolbox for handling decomposers and definable cliquewidth, and at some point the Binary Lemma will easily follow from the already gathered observations.

\subsection{Toolbox for decomposers}\label{sec:toolbox}

We now give several useful tools for handling decomposers.

\subsubsection*{Filtering and Transfering Structure.}
In this section, we establish two simple lemmas which crucially rely on decomposers being
\emph{origin-preserving}, that is, satisfying condition \ref{c:origin} of
Definition~\ref{def:decomposer}. In the following, let $\{E\}$ be the vocabulary of
graphs, where $E$ is the binary adjacency relation, and let $\Delta_k$ be the vocabulary of width-$k$ cliquewidth
decompositions. We assume that $E\notin \Delta_k$.

The first of our lemmas allows us to make sure that a
nondeterministic transduction that is supposed to be a decomposer is
correct by filtering out outputs that are not cliquewidth decompositions of
the input graph. Recall that for every input-output pair $(G,t)$ of a decomposer
$\mathcal D$ the structure $t$ is a cliquewidth decomposition of $G$ and
the origin mapping of $\mathcal D$ induces an isomorphism from $G_t$
to $G$. In general, for an \mso-transduction $\mathcal{I}$ from $\{E\}$
to $\Delta_k$, we say that $\mathcal{I}$ \emph{decomposes} a graph $G$ if there is a some output $t$ of $\mathcal{I}$ on input $G$ such that $t$ is a cliquewidth decomposition of $G$ 
and the origin mapping of $\mathcal{I}$ induces an isomorphism from $G_t$ to $G$.

\begin{lem}[Filter Lemma]
  Let $\mathcal{I}$ be an \mso-transduction with input vocabulary being the vocabulary of graphs $\{E\}$ and output vocabulary $\Delta_k$. Then there is
  a width-$k$ decomposer $\mathcal D$ that decomposes the same graphs as~$\mathcal{I}$.
\end{lem}
\begin{proof}
  Let $R$ be a binary relation symbol not contained in $\{E\}\cup\Delta_k$.
  By copying the input, we can modify $\mathcal{I}$ to obtain a
  transduction $\mathcal{I}'$ from $\{E\}$ to $\{E,R\}\cup\Delta_k$ that for every input-output pair $(G,t)$
  of $\mathcal{I}$ has an input-output pair $(G,A)$, where $A$ is the
  structure obtained from the disjoint union of $G$ and $t$ by adding
  a binary relation $R$ that connects all elements with the same origin.
  Recall that one of atomic transductions is {\em{filtering}}, which amounts to discarding all structures not satisfying a fixed \mso sentence.
  Using this mechanism, we can discard those pairs $(G,A)$ where
  the underlying $t$ is not a cliquewidth decomposition of $G$ for which the
  origin mapping induces an isomorphism from $G_t$ to $G$. (Note that
  we cannot check whether $G_t$ is isomorphic to $G$, but we can check
  whether the origin mapping is an isomorphism. This is the main reason
  why we require decomposers to be origin preserving.) Finally,
  we can restrict the universe of $A$ to retrieve the original $t$.
\end{proof}

The Filter Lemma implies that to prove Theorem~\ref{thm:main}, it
suffices to prove that for every $k$ there is an $\ell$ and an \mso-transduction
from $\{E\}$ to $\Delta_\ell$ that decomposes
  all graphs of linear cliquewidth $k$.

The second consequence of the decomposers being origin-preserving is
that we can transfer additional structure present on the input graphs to the
graphs constructed by the output decompositions. For vocabularies
$\Sigma\subseteq\Sigma'$, the \emph{$\Sigma$-reduct} of
a $\Sigma'$-structure $A'$ is the $\Sigma$-structure $A$ that has the same
universe as $A'$ and coincides with $A'$ on all relations in $\Sigma$. Conversely, a \emph{$\Sigma'$-expansion} of
a $\Sigma$-structure $A$ is the $\Sigma'$-structure $A'$ such that $A$
is the $\Sigma$-reduct of $A'$.

\begin{lem}[Transfer Lemma]
  Let $\mathcal D$ be a width-$k$ decomposer and let $\Sigma$ be a
  vocabulary disjoint from $\{E\}\cup\Delta_k$. Then there is
  an \mso-transduction $\mathcal D^\star$ from $\{E\}\cup\Sigma$ to
  $\Delta_k\cup\{E\}\cup\Sigma$ such that for every input-output pair $(G,t)$
  of $\mathcal D$
  and every $\{E\}\cup\Sigma$-expansion $G^\star$ of $G$ there is a unique
  $\Delta_k\cup\{E\}\cup\Sigma$-expansion $t^\star$ of $t$ such that
  $(G^\star,t^\star)$ is an input-output pair of $\mathcal D^\star$ and the origin
  mapping restricted to the leaves of $t$ is an isomorphism
  from the induced substructure of $t^\star$ to $G^\star$. 
\end{lem}
\begin{proof}
  Recall that $\mathcal D$, being a width-$k$ decomposer, is a sequence of atomic transductions with input vocabulary $\{E\}$ and output vocabulary $\{E\}\cup \Delta_k$.
  We apply exactly the same sequence of atomic transductions, except that all the additional relations from $\Sigma$ are always kept intact.
  The claim follows by the assumption that $\mathcal D$ is origin-preserving (condition \ref{c:origin} of Definition~\ref{def:decomposer}).
\end{proof}

We will use this lemma to transfer colors of the input graph of a
decomposer to the~output.

\subsubsection*{Enforcing a fixed partition.}
Given a cliquewidth decomposition of a graph, say of width $k$, and a partition of the vertex set into $p$ subsets, which may be non-related to the decomposition,
one can adjust the decomposition at the cost of using $k\cdot p$ colors instead of $k$ so that the final color partition of the decomposition matches the given one. 
Informally, this can be done by just enriching each original label
with information to which subset of the final partition a vertex belongs. The following general-usage lemma formalizes this, 
and shows that the transformation may be performed by means of an \mso transduction

\begin{lem}[Color Enforcement Lemma]\label{lem:enrich}
For every $k,p\in \Nats$, there exists a deterministic
\mso transduction $\mathcal{E}_{k,p}$ with the following properties. 
The input vocabulary of $\mathcal{E}_{k,p}$ is the vocabulary of cliquewidth decompositions of width $k$ with leaves colored by $p$ unary predicates.
The output vocabulary is the vocabulary of cliquewidth decomposition of width $k\cdot p$.
Finally, on an input decomposition $t$ with leaves partitioned into $(V_1,\ldots,V_p)$ using the unary predicates, 
the output of $\mathcal{E}_{k,p}$ is a decomposition $t'$ of the same graph, where in the result of $t'$ the color of each vertex from $V_i$ is equal to $i$, for all $i\in [p]$.
\end{lem}
\begin{proof}
The decomposition $t$ is first adjusted to a decomposition $t''$ of width $k\cdot p$ with the following property: in the result of $t''$,
the final color of every vertex is a pair consisting of its color in the result of $t$ and the index $i$ such that the leaf corresponding to the vertex belongs to $V_i$.
This correction can be made by leaving the shape of $t$ intact, and performing a straightforward modification to the labels of nodes.
For instance, for a Join node, whenever the original label in $t$ requested adding edges between colors $c$ and $d$, the new label in $t''$ requests adding edges
between colors $(c,i)$ and $(d,j)$ for all $i,j\in [p]$. Finally, we obtain $t'$ by adding a recoloring step on top of $t''$ that removes the first coordinate of every color.
\end{proof}

Using the Color Enforcement Lemma, we can give a proof of the Binary Lemma.

\begin{proof}[Proof of the Binary Lemma]
Let $m=\max(\dcw(\sigma),\dcw(\tau))$, and let $\mathcal{D}_\sigma$ and $\mathcal{D}_\tau$ be decomposers of size at most $m$ such that $\mathcal{D}_\sigma$ produces at least one
output on the underlying graph of $\sigma$, and similarly for $\mathcal{D}_\tau$. 
Using coloring, we first guess the partition of the vertex set into vertices that belong to the underlying graphs of $\sigma$ and $\tau$.
Next, we guess the color partition in the underlying graph of $\sigma$.
Finally, for the underlying graph of $\tau$, we guess the partition of its vertices according to profiles in $\tau$.
Note that the validity of this guess, or more precisely the fact that the adjacency between the $\sigma$-part and the $\tau$-part depends only on the (color,profile) pair of respective vertices,
can be checked using a filtering step.

We now apply $\mathcal{D}_\sigma$ to the $\sigma$-part of the graph, yielding a cliquewidth decomposition of the underlying graph of $\sigma$ of width at most $m$.
By applying the transduction $\mathcal{E}_{m,k}$ given by the Color Enforcement Lemma to this decomposition, by the Transfer Lemma
we can assume that the result of the obtained decomposition $t_\sigma$ has the color partition
equal to the color partition of $\sigma$. 
Similarly, by applying $\mathcal{D}_\tau$ followed by $\mathcal{E}_{m,2^k}$ for the profile partition, we turn the $\tau$-part of the graph into its cliquewidth decomposition $t_\tau$ whose result has the
color partition equal to the profile partition in $\tau$. 
Since the adjacency between the $\sigma$-part and the $\tau$-part depends only on the (color,profile) pair of respective vertices, it now suffices to add one binary Join node, with the 
roots of $t_\sigma$ and $t_\tau$ as children, where we request adding edges between appropriate pairs of vertices, selected by color on the $\sigma$-side and profile on the $\tau$-side.
\end{proof}

\subsubsection*{Combining many decomposers.}
In the setting of the Idempotent Lemma, the graph consists of multiple pieces, each having small definable cliquewidth. 
Thus, we may think that for each piece we have already constructed a decomposer (w.l.o.g. the same one, as we can take the union of the input decomposers), 
and now we need to put all these decomposers together. In particular, we will need to apply the decomposers ``in parallel''
to all the considered pieces. The following Parallel Application Lemma formalizes this idea.

For a vocabulary $\Sigma$ and a sequence $\Af_1,\ldots,\Af_n$ of 
$\Sigma$-structures, the {\em{disjoint union}} of the structures $\Af_1,\ldots,\Af_n$, denoted $\biguplus_{i=1}^n \Af_i$, is the structure over vocabulary $\Sigma\cup \{\sim\}$, where $\sim$ is a binary symbol, defined as follows:
\begin{itemize}
\item the universe of $\biguplus_{i=1}^n \Af_i$ is the disjoint union of the universes of $\Af_i$ for $i\in [n]$;
\item for each symbol $R\in \Sigma$, the interpretation of $R$ in $\biguplus_{i=1}^n \Af_i$ is the union of its interpretations in structures $\Af_i$ for $i\in [n]$;
\item $\sim$ is interpreted as the equivalence relation on the universe of $\biguplus_{i=1}^n \Af_i$ that relates pairs of elements originating in the same structure $\Af_i$.
\end{itemize}

\begin{lem}[Parallel Application Lemma]\label{lem:parallel}
Let $\mathcal{I}$ be an \mso transduction with input vocabulary $\Sigma$ and output vocabulary $\Gamma$.
Then there is an \mso transduction $\widehat{\mathcal{I}}$ with input vocabulary $\Sigma\cup \{\sim\}$, output vocabulary $\Gamma\cup \{\sim\}$, and the following semantics:
for every sequence $(\Af_1,\Bf_1),\ldots,(\Af_n,\Bf_n)$ of pairs of $\Sigma$- and $\Gamma$-structures,
we have
$(\biguplus_{i=1}^n \Af_i,\biguplus_{i=1}^n \Bf_i)\in \widehat{\mathcal{I}}$
if and only if $(\Af_i,\Bf_i)\in \mathcal{I}$ for all $i\in [n]$.
\end{lem}
\begin{proof}
Observe that it suffices to verify the lemma on atomic transductions.
For copying and coloring the claim is trivial: we can take the same operation.
For universe restriction, say using an \mso predicate $\psi(x)$, we use universe restriction using a predicate $\psi'(x)$ that is constructed from $\psi(x)$ by relativizing it to
the $\sim$-equivalence class of $x$, that is, adding a guard to every quantifier that restricts its range to (sets of) elements $\sim$-equivalent to $x$.
For interpretation, we similarly modify each \mso formula $\varphi_R(x_1,\ldots,x_r)$ by additionally requiring that the elements $x_1,\ldots,x_r$ are pairwise $\sim$-equivalent, and relativizing
the formula to the $\sim$-equivalence class of $x_1,\ldots,x_r$.
Finally, for filtering, say using an \mso sentence $\psi$, we use filtering using an \mso sentence saying that for every equivalence class of $\sim$, the formula $\psi$ relativized to this equivalence class holds.
\end{proof}

We now proceed to the final tool for decomposers: the Combiner Lemma. 
In principle, it formalizes the idea that in the setting of the Idempotent Lemma, having defined the block order (roughly, using the Definable Order Lemma), 
we may construct cliquewidth decompositions for individual pieces (derivations), obtained by applying small decomposers in parallel, into a cliquewidth decomposition of the whole graph. 

Define an \emph{order-using decomposer} to be an \mso transduction which inputs a graph $G$ together with a linear quasi-order on its vertices and which outputs cliquewidth decompositions of the input graph. 
On a given input, an order-using decomposer might produce several outputs, possibly zero.

\begin{lem}[Combiner Lemma]\label{lem:combiner} 
For every $m  \in \Nat$ there is an order-using decomposer $\mathcal{D}$ with the following property. 
Let  $\tau_1,\ldots,\tau_n$ be $k$-derivations whose underlying graphs have definable cliquewidth at most $m$. 
Let $G$ be the underlying graph of $\tau_1 \cdots \tau_n$ and $\bkleq$ be the block order arising from decomposition $\tau_1\cdots \tau_n$. 
Then $\mathcal{D}$ produces at least one output on~$(G,\bkleq)$.
\end{lem}
\begin{proof}
In the following, we describe the order-using decomposer $\mathcal{D}$.
First, using coloring guess the partition of the vertex set into sets $\{U_c\colon c\in \Cells\}$ such that $U_c=\bigcup_{i\in [n]}\tau_i[c]$.
Then the cell $\tau_i[c]$ may be recovered as the intersection of $U_c$ with the underlying graph of $\tau_i$, which in turn can be identified as a single equivalence class of the block equivalence.

By assumption, for each $i\in [n]$ there is a decomposer $\mathcal{D}_i$ of size at most $m$ that applied to the underlying graph of $\tau_i$ produces at least one output.
By the Color Enforcement Lemma applied to the cell partition $\{\tau_i[c]\colon c\in \Cells\}$ and the Transfer Lemma, we may assume that the result of each 
$\mathcal{D}_i$ has color partition coinciding with this cell partition.
After this operation, the sizes of all decomposers $\mathcal{D}_i$ are still bounded by some $m'$ depending only on $m$ and $k$.

Let now $\mathcal{J}$ be the union of all decomposers of size at most $m'$.
As we argued before, $\mathcal{J}$ is a decomposer of size bounded by a function of $m'$ such that $\mathcal{J}$ applied to the underlying graph of any $\tau_i$ has at least one output. Moreover, this output is
a cliquewidth decomposition $t_i$ of the underlying graph of $\tau_i$ such that in the result of $t_i$, the color partition is equal to the cell partition in $\tau_i$. 

Let $\bkeq$ be the block equivalence in $\tau_1\cdots\tau_n$, interpreted from the block order $\bkleq$.
Apply Parallel Application Lemma to $\mathcal{J}$ and $\bkeq$, yielding an \mso transduction $\widehat{\mathcal{J}}$ that, when applied to the whole structure,
turns the underlying graph of each $\tau_i$ into its cliquewidth decomposition $t_i$ as above.
Since each $t_i$ originates in vertices of the underlying graph of $\tau_i$, on decompositions $t_i$ we still have the order $\bkleq$ present in the structure.

It now remains to combine decompositions $t_i$ sequentially. 
We do it as follows.
For every $i\in [n-1]$ we create two nodes $a_i,b_i$, for instance by copying the roots of the decompositions $t_i$ for $i\in [n-1]$ two times.
Then we connect these nodes into a path, called the {\em{spine}}, so that each $a_i$ is a child of $b_i$, and each $b_i$
is a child of $a_{i+1}$ (except $i=n-1$). It is easy to do it in a single interpretation step, as the order $\bkleq$ is present in the structure, 
so for every decomposition $t_i$ we can interpret the next decomposition $t_{i+1}$.
Further, we make the root of $t_{i+1}$ a child of $a_i$ for each $i\in [n-1]$, and moreover the root of $t_1$ becomes a child of $a_1$; again, this can be done in one interpretation step. 
This establishes the
shape of the final decomposition, where $b_{n-1}$ is the root. 

For the labels of nodes, each $a_i$ is labeled by a Join operation, and each $b_i$ is labeled by a Recolor operation. 
In the colored graphs computed along the spine, the consecutive colors assigned to every vertex, say originating from $\tau_i$, are equal to the color that 
would be assigned to this vertex in derivations $\tau_i$, $\tau_i\cdot \tau_{i+1}$, $\tau_{i}\cdot \tau_{i+1}\cdot \tau_{i+2}$, and so on.

The Join operation at node $a_i$ requests adding edges between every vertex $u$ coming from the child on the spine ($b_{i-1}$, or $t_1$ if $i=1$), 
and every vertex $v$ coming from decomposition $t_{i+1}$, whenever the color of $u$ belongs to the profile of $v$. 
Recall here that the color partition in the result of $t_{i+1}$ matches the cell partition in $\tau_{i+1}$, so the profiles in $\tau_{i+1}$ are encoded in the colors in the result of $t_{i+1}$.
Also, we may assume that the colors originating from the subtree below $b_{i-1}$ are pairwise different than colors originating from $t_{i+1}$, so in the Join at $a_i$ no two colors are merged.

The Recoloring operation at node $b_i$ removes the information about profiles from the colors of
vertices originating from $t_{i+1}$, and adjusts colors for vertices coming from below the spine (i.e., originating in decompositions $t_{i'}$ for $i'\leq i$)
according to the recoloring applied in $\tau_{i+1}$. Observe that for each node $b_i$ we may
guess this recoloring nondeterministically, by guessing, for every function $\phi\colon [k]\to [k]$, a unary predicate that selects nodes $b_i$ where recoloring $\phi$ should be used.
By appealing to the Filter Lemma, we can always check that in the end 
we have indeed obtained a cliquewidth decomposition of the input graph.
 Hence, even though some of the nondeterministic guesses may lead to constructing a cliquewidth decomposition whose result is different from the input graph,
these guesses will be filtered out at the end.
\end{proof}

\subsection{Definable cliquewidth under restriction of the universe}

In the Idempotent Lemma we assume that each individual $k$-derivation $\sigma_i$ has bounded definable cliquewidth, say by $K$, which means that we have a decomposer of size at most $K$
that constructs a cliquewidth decomposition of the underlying graph $G_i$ of $\sigma_i$. 
However, recall that the Definable Order Lemma does not provide the full block order (that could be fed to the Combiner Lemma), only its restriction to the connected components of some $Z$-flip of the graph.
Therefore, the graphs $G_i$ are not directly available to the transduction; we are able to construct only their restrictions to the connected components of said $Z$-flip.

It would be now convenient to claim that definable cliquewidth is closed under taking induced subgraphs, similarly as the standard cliquewidth is, so that the Combiner Lemma could be applied to each
connected component of the $Z$-flip separately.
This, however, is not immediate, as the decomposer for the induced subgraph would need to work only on this induced subgraph.
In fact, we do not know whether this statement is true at all, but we can prove a weaker variant that turns out to be sufficient for our needs.

Let $G$ be an undirected graph and let $(V_0,V_1)$ be a partition of the vertex set of $G$ into two sets.
We define the \emph{rank} of the partition $(V_0,V_1)$ as the number of equivalence classes in the following equivalence relation on vertices. 
Two vertices $v,w$ are considered equivalent if they both belong to the same $V_i$ for some $i \in \set{0,1}$, and the sets of neighbors of $v$ and $w$ within  $V_{1-i}$ are the same. 
Define the \emph{rank} of an induced subgraph $H$ of $G$, denoted $\rank{G,H}$, to be the rank of the partition $(V(H),V(G)\setminus V(H))$ in $G$.
We prove that the definable cliquewidth of an induced subgraph is bounded by a function of the definable cliquewidth of the larger graph, provided the rank of the induced subgraph within the larger graph is bounded. 

\begin{lem}\label{lem:induced-dcw}
There is a function $f \colon \Nat \to \Nat$ such that for every graph $G$ and an induced subgraph~$H$ of $G$, we have
\begin{align*}
  \dcw(H) \le f (\max(\dcw(G), \rank{G,H})).
\end{align*}
\end{lem}

Before we proceed to the proof of Lemma~\ref{lem:induced-dcw}, we give the following lemma about replacing a part of a graph subject to preserving the satisfaction of an \mso sentence.
Its proof is completely standard, but we give it for completeness.

\begin{lem}[MSO Pumping Lemma]\label{lem:mso-localisation}
Let $\varphi$ be an \mso sentence over  graphs (i.e.~the universe is the vertex set and the vocabulary consists of one binary edge predicate). 
Let $G$ be a  graph that satisfies $\varphi$. For every  induced subgraph $H$ of $G$, there is some $G'$ such that:
\begin{enumerate}[(1)]
\item\label{i1} $G'$ satisfies $\varphi$;
\item\label{i2} $H$ is an induced subgraph of $G'$; and
\item\label{i3} the number of vertices in $G'-H$ is bounded by a constant depending only on $\varphi$ and $\rank{G,H}$.
\end{enumerate}
\end{lem}
\begin{proof}
Define $\mathcal{N}$ to be the family of neighborhoods in $H$ of vertices from $G-H$, i.e.
\begin{align*}
\mathcal{N} \eqdef  \set{ \set{v  : \mbox{$v$ is a vertex in $H$ adjacent to $w$}} : w \mbox{ is a vertex in } G-H}.
\end{align*}
Observe that if the rank of $H$ in $G$ is $r$, then $|\mathcal{N}|\leq r$.
This is because every neighborhood in $\mathcal{N}$ is the union of a collection of equivalence classes of the equivalence relation considered in the definition of $\rank{G,H}$.  

Define $F$ to be the $\mathcal{N}$-colored graph obtained from $G-H$ by coloring each vertex by its neighborhood in $H$. 
We treat $F$ as a relational structure with one binary edge predicate and $|\mathcal{N}|$ unary predicates, one for each color.
Let $q$ be the quantifier rank of $\varphi$. 
Choose $F'$ to be the smallest $\mathcal{N}$-colored graph which satisfies the same \mso sentences of quantifier rank $q$ as~$F$.
The size of $F'$ is bounded by a constant depending on $|\mathcal{N}|$ and $q$, which in turn depend only on $\rank{G,H}$ and $\varphi$. 
Define $G'$ to be the following graph: 
we take the disjoint union of $F'$ and~$H$, forget the coloring in $F'$, and then for each vertex $v$ in $F'$, 
we connect it to  those vertices in $H$ which were in its (now forgotten) color. 
Using an Ehrenfeucht-Fraisse argument, it is straightforward to show that $G'$ and $G$ satisfy the same \mso sentences of quantifier rank at most $q$.
In particular, $G'$ satisfies $\varphi$.
\end{proof}

We would like to remark that the number of vertices in $G'-H$ in the above lemma can be computed given: $\varphi$, the rank $\rank{G,H}$ and the cliquewidth of $G$. 
(Note that the lemma asserts a stronger property, namely that the number of vertices in $G'-H$ is bounded only by $\varphi$ and the rank $\rank{G,H}$, and there is no dependency on the cliquewidth of $G$. 
Nevertheless, for computability we also use the cliquewidth of $G$, and this additional dependency is not an issue for our intended application in the proof of Lemma~\ref{lem:induced-dcw}, 
where we have an upper bound on the cliquewidth of $G$ anyway.) Indeed, given numbers $r,k,k' \in \set{1,2,\ldots}$ and an \mso formula $\varphi$,  consider the following statement:
\medskip
\begin{quote}
	($\star$) For every graph $G$ of cliquewidth at most $ k$, and every induced subgraph $H$ of $G$ satisfying
	\begin{align*}
\rank{G,H} < r,
\end{align*}
	 there exists a graph $G'$ which satisfies items \ref{i1} and \ref{i2} from Lemma~\ref{lem:mso-localisation} and such that the number of vertices in $G'-H$ is at most $k'$.  
\end{quote}
\medskip
Lemma~\ref{lem:mso-localisation} implies that for every $\varphi, r,k$ there exists some $k'$ which makes ($\star$) true. 
By the following lemma, to compute (given $\varphi,r,k$) the smallest $k'$ which makes ($\star$) true we can go through each candidate for $k'$ and check whether ($\star$) is satisfied. 
Summing up, the bound on the size of $G'-H$ in item \ref{i3} of Lemma~\ref{lem:mso-localisation} is computable, assuming additionally that we know the cliquewidth of $G$.

\begin{lem}
  For every $\varphi,r,k,k'$ one can decide whether ($\star$) holds.
\end{lem}
\begin{proof}
For an \mso formula $\varphi$ and  $k' \in \set{1,2,\ldots}$, consider the following property of graphs:

  {\small
\begin{align}\label{eq:additional-k}
\set{H \colon \text{there is some $G' \models \varphi$ such that} \begin{cases}
	\text{$G'$ satisfies $\varphi$,}\\ \text{$H$ is an induced subgraph of $G'$, and}\\ \text{the number of vertices in $G'-H$ is at most $k'$}
\end{cases}}	
\end{align}
  }

\noindent It is straightforward to see that the above property is also \mso definable. Indeed, one may existentially guess the isomorphism class of $G'-H$ and its adjacencies to $H$, because $G'-H$ has bounded size,
and then rewrite $\varphi$ by simulating quantification over the additional vertices in $G'-H$ within syntax.
 The property  ($\star$) asks if the formula expressing~\eqref{eq:additional-k} is true for all graphs of cliquewidth $\le k$.  
 Checking whether an \mso formula is true for all graphs of cliquewidth $\le k$ is a decidable problem, see~\cite[Section 7.5]{0030804}.
\end{proof}

\begin{proof}[Proof of Lemma~\ref{lem:induced-dcw}]
Let $\ell=\max(\dcw(G),\rank{G,H})$.
Our goal is to show that the definable cliquewidth of $H$ is bounded by a function of $\ell$. 
By definition of definable cliquewidth, there is a decomposer  $\mathcal{D}$ which  produces at least one output on $G$. 
From the Backwards Translation Theorem applied to the sentence ``true'' it follows that the domain of the decomposer $\mathcal{D}$, i.e.~those graphs where it produces at least one output, 
is \mso definable, say by a sentence $\varphi$. 
Apply the \mso Pumping Lemma to $\varphi$ and the graph $H \subseteq G$, yielding 
a graph $G'\supseteq H$ in the domain of $\mathcal{D}$ such that $H$ is an induced subgraph of $G'$ and the number of vertices in $G'-H$ is bounded by a constant $m$ depending only on $\ell$. By the discussion after the \mso Pumping Lemma, the constant $m$ can be effectively computed.

Since $G'$ is in the domain of $\mathcal{D}$, we have that $\mathcal{D}$ applied on $G'$ produces at least one output, say $t$.
By the definition of a decomposer, $t$ is a cliquewidth decomposition of $G'$.
Then, a cliquewidth decomposition of $H$ can be obtained from $G'$ by removing all leaves of $t$ corresponding to the vertices of $G'-H$, and performing straightforward cleaning operations.

We now construct a decomposer for $H$ as follows.
First, we copy any vertex of the graph $m$ times, and using coloring and interpretation we (nondeterministically) turn $H$ into $G'$.
Then, we apply $\mathcal{D}$ to $G'$, yielding some cliquewidth decomposition $t$ of $G'$.
By the Transfer Lemma, we can assume that the original relations are preserved on the leaves of~$t$.
Therefore, we can now remove the leaves of $t$ corresponding to vertices of $G'-H$ and perform the clean-up operations; it is easy to see that this can be done by means of an \mso transduction.
\end{proof}

As argued, the computability of the bound in item \ref{i3} of Lemma~\ref{lem:mso-localisation} entails the computability of the function~$f$ provided by Lemma~\ref{lem:mso-localisation}.

\subsection{Completing the proof of the Idempotent Lemma}
\label{sec:use-dol}

With all the tools prepared, we complete the proof of the Idempotent Lemma.

\begin{proof}[Proof of the Idempotent Lemma] 
Let
$\sigma_1,\ldots,\sigma_n$ be $k$-derivations as in the Idempotent Lemma, i.e., with the same idempotent abstraction~$e$.
Let 
\[K=\max_{i\in [n]}\,\dcw(\sigma_i).\]
Write $\sigma=\sigma_1 \cdots \sigma_n$. Let $G$ be the underlying graph of $\sigma$, let $G_i$ be the underlying graph of $\sigma_i$ for each $i\in [n]$, and let $\bkleq$ be the block order in $G$ compliant
with the decomposition $\sigma_1 \cdots \sigma_n$. 
It suffices to describe a decomposer of size bounded in terms of $k$ and $K$ that constructs a cliquewidth decomposition of $G$.

First, using coloring we enrich the structure with unary predicates that encode the partition of the vertex set of $G$ into cells $\sigma[c]$, for $c\in \Cells$.
Then we apply the Definable Order Lemma to $\sigma_1,\ldots,\sigma_n$, yielding $Z$ and $\sim$.
Note that $Z$ is chosen among $2^{|\binom{\Cells}{1,2}|}$ options, so we can nondeterministically guess $Z$ using, say, some coloring.
Having $Z$ fixed, the equivalence relation $\sim$ (being in the same connected component of the $Z$-flip of $\sigma$) can be added to the structure using interpretation.
By the Definable Order Lemma, we can add also the relation $\bkleq\cap \sim$ to the structure, as this increases the size of the transduction only by a function of $k$ and $K$.

The next claim says that restricting blocks to equivalence classes of $\sim$ yields graphs of bounded definable cliquewidth. 
Here, we will crucially use the results from the last section on how definable cliquewidth behaves
under restricting the vertex set.

\begin{clm}\label{claim:subgraph}
There is $m\in \Nats$ depending only $k$ and $K$ such that for every equivalence class $F$ of $\sim$ and every $i \in [n]$, 
the subgraph induced in $G_i$ by vertices contained in $F$ has definable cliquewidth at most~$m$.
\end{clm}
\begin{proof}
Take any $i\in [n]$ and any connected component $F$ in the $Z$-flip of $\sigma$. 
Let $H_i=G_i[V(F)\cap V(G_i)]$, that is, $H_i$ is the subgraph induced in $G_i$ by vertices contained in $F$. 
Observe that the rank of $H_i$ inside $G_i$ is bounded by $2k\cdot 2^k$, i.e., twice the number of cells.
Indeed, since $Z$-flip changes only the adjacency between whole cells, and after performing the $Z$-flip there are no edges between the vertices of $H_i$ and the rest of vertices of $G_i$,
for every cell $c\in \Cells$ we have that both $(\sigma[c]\cap V(G_i))\cap V(F)$ and $(\sigma[c]\cap V(G_i))\setminus V(F)$ are contained in the same equivalence
class of the equivalence relation considered in the definition of the rank of $H_i$ inside $G_i$.
The claim now follows directly from Lemma~\ref{lem:induced-dcw}.
\end{proof}


Now apply the Combiner Lemma to the parameter $m$ given by Claim~\ref{claim:subgraph}, yielding an order-using decomposer $\mathcal{D}$ satisfying the following.

\begin{clm}\label{cl:it-is-too-late}
For each equivalence class $F$ of $\sim$, the order-using decomposer $\mathcal{D}$ produces at least one output on $(G,\bkleq)$ restricted to the vertices of $F$.
\end{clm}

Recall that the cell partition $\{\sigma[c]\colon c\in \Cells\}$ has been guessed and added to the structure via unary predicates.
By the Transfer Lemma, we may assume that this cell partition is preserved on the leaves of the output $t_F$ of applying $\mathcal{D}$ to an equivalence class $F$ of $\sim$.
Hence, by appending an appropriate transduction given by the Color Enforcement Lemma for the cell partition, 
we may assume without loss of generality that the color partition in the result of $t_F$ is equal to the restriction of the cell partition to $F$.

Let $\widehat{\mathcal{D}}$ be the \mso transduction given by the Parallel Application Lemma for $\mathcal{D}$.
Using $\sim$ as the input equivalence relation for $\widehat{\mathcal{D}}$, we infer that $\widehat{\mathcal{D}}$ applied to the current structure 
turns every equivalence class $F$ of $\sim$ into a cliquewidth decomposition $t_F$ of the subgraph induced by $F$ in $G$.
Moreover, the color partition in the result of $t_F$ is equal to the restriction of the cell partition to $F$.

It now suffices to add a new root node and attach all the root nodes of decompositions $t_F$ as its children.
This new root node is labeled by a Join operation, where we request that for all $c,d\in \Cells$ for which $\{c,d\}\in Z$, 
an edge should be added between every pair of vertices $u,v$ such that $u$ belongs to cell $c$, $v$ belongs to cell $d$, and $u,v$ originate from
different decompositions $t^F$. Note that this is possible since in the result of each decomposition $t^F$, the color partition matches the appropriate restriction of the cell partition.
Since the connected components of the $Z$-flip of $G$ are pairwise non-adjacent in this $Z$-flip by definition, it is clear that the structure constructed in this manner is a cliquewidth decomposition of $G$.
\end{proof}

We remark that the operation performed in the final paragraph of the proof of the Idempotent Lemma is the only place in the argumentation where we apply the Join operation over an unbounded number of arguments. Consequently, this is the only moment in the proof where we inherently use that we are allowed to construct a general cliquewidth decomposition, and not necessarily a linear cliquewidth decomposition.

\section{Proof of the Definable Order Lemma}\label{app:def-ord}

In this section we present the proof of the Definable Order Lemma. Let
$\sigma_1,\ldots,\sigma_n$ be $k$-derivations which have the same
idempotent abstraction. Recall that our goal is to prove that there is
a set $Z\subseteq \binom{\Cells}{1,2}$ such that in the $Z$-flip of
$\sigma_1 \cdots \sigma_n$ the relation $\sim \cap \preceq$ has
interpretation complexity over $G$ bounded by a function of $k$.

The proof is rather lengthy and consists of several technical steps. To facilitate the understanding, we divided it into several subsections. Each subsection starts with a semi-formal description of the current goal and its intuitive meaning.

\subsection{Setup}

Before we proceed with the proof, let us introduce some basic definitions.

\subsubsection*{Block order on (pairs of) cells.}
For a cell $c\in \Cells$, by $U_c$ we denote the set of vertices of $G$ comprising all vertices from $c$-cells 
in respective derivations $\sigma_s$.
In other words,
\[U_c\eqdef\bigcup_{s\in [n]} \sigma_s[c].\]
Let us stress that $U_c$ may be different from $\sigma[c]$, that is, the set of vertices from the $c$-cell in the overall derivation $\sigma$.
This is because the profile and the color of a vertex in $\sigma$ may differ from its profile and color in respective $\sigma_s$.

By $\bkleq_c$ and $\bkeq_c$ we denote the restriction of $\bkleq$ and $\bkeq$ to $U_c$, respectively.
Moreover, for cells $c,d\in \Cells$, by $\bkleq_{c,d}$ and $\bkeq_{c,d}$ we denote the restriction of $\bkleq$ and $\bkeq$ to pairs from $U_c\times U_d$, respectively. 
Our first goal is to interpret the above relations for as large subset of cells (resp. pairs of cells) as possible. 

\subsubsection*{Moduli.}
For distinguishing neighboring blocks we introduce the following definitions.
By {\em{moduli}} we mean the remainders modulo $7$, that is, the elements of the set $\{0,1,\ldots,6\}$.
All arithmetic on moduli is performed modulo $7$.
The {\em{distance}} between two moduli $a,b$, denoted $\dist(a,b)$, is the smaller of the numbers $(a-b)\bmod 7$ and $(b-a)\bmod 7$.
Moduli $a,b$ are {\em{neighboring}} if the distance between them is $1$.
Let the {\em{modulus}} of the $i$-th block be equal to $i\bmod 7$, for each $i\in [n]$. 
The modulus of a vertex $u$ of $G$, denoted $\modulus{u}$, is the modulus of the block to which it belongs.
For a modulus $a\in \{0,1,\ldots,6\}$, by $W_a$ we denote the set of all vertices of $G$ that have modulus $a$.

We can now define a structure $\Ga$ that is the expansion of $G$ with the following unary predicates:
\begin{itemize}
\item for each $c\in \Cells$, a unary predicate that selects the vertices of $U_c$; and
\item for each modulus $a$, a unary predicate that selects the vertices of $W_a$.
\end{itemize}
The reader should think of these unary predicates as of some auxiliary information that is helpful in analyzing the graph for the purpose of interpreting the block order.
These unary predicates will be (existentially guessed) monadic parameters $X_i$ from the definition of interpretation complexity, hence we may simply assume that we work over the structure $\Ga$.

\subsubsection*{Idempotent recolorings.}
Observe that since the common abstraction $e=(L,\rho,\phi)$ is idempotent in $T_k$, 
all the input derivations $\sigma_s$ have to have the same recoloring $\phi$, and this recoloring has to be idempotent in the semigroup of functions from $[k]$ to $[k]$, endowed with the composition operation. 
This is because when composing two $k$-derivations, we compose their recolorings.
It is easy to see that a function $\psi\colon [k]\to [k]$ is idempotent if and only if each $i$ belonging to the image of $\psi$ is a fixed point of $\psi$, that is, $\psi(i)=i$ for each $i\in \psi([k])$.
Thus we have $\phi(i)=i$ for each $i\in \phi([k])$.

It is instructive to consider what this means in terms of recolorings, when each $\sigma_j$ is treated as a sequence of instructions. 
Suppose that some vertex $u$ belongs to the underlying graph of $\sigma_s$, for some $s\in [n]$, and its color in $\sigma_s$ is $i$.
This means that after applying all the operations in $\sigma_s$, the color of $u$ is $i$, however this color may further change due to recolorings applied in each $\sigma_{t}$ for $t>s$.
For instance, the application of the recoloring $\phi$ in $\sigma_{s+1}$ changes the color of $u$ from $i$ to $\phi(i)$. However, since $\phi(\phi(i))=\phi(i)$, the application of
recoloring $\phi$ in every further $\sigma_{t}$, i.e. for $t\geq s+2$, will not change the color of $u$, and this color will stay equal to $\phi(i)$ up to the end of the sequence. 

\subsection{Types of pairs of cells}\label{sec:types}

The first step is to examine all pairs of essential cells $(c,d)$ and partition them into {\em{types}}; recall here that a cell $c$ is {\em{essential}} if $\sigma_s[c]$ is nonempty for every $s$. This partition depends on the recoloring $\phi$ and is formally defined in terms of colors and profiles of $c$ and $d$, but the intuition is as follows. Consider two blocks $G_s$ and $G_t$ that are far from each other: $|s-t|>1$. Pick any $u\in \sigma_s[c]$ and $v\in \sigma_t[d]$. Assume for a moment that $s<t$. By the idempotency of the recoloring $\phi$, it is easy to see that the cells $c$ and $d$ uniquely determine whether $u$ and $v$ are adjacent: this depends on the color of $c$ and the profile of $d$. Similarly, if $t<s$ then $c$ and $d$ also uniquely determine whether $u$ and $v$ are adjacent. Thus, with the pair $(c,d)$ we can associate two boolean values: the outcomes of the checks described above. If both these values are false, then the pair $(c,d)$ shall be called {\em{negative}}: pairs from $U_c\times U_d$ that are far in the block order are always non-adjacent. If both are true, then the pair is {\em{positive}}: pairs from $U_c\times U_d$ that are far in the block order are always adjacent. The interesting situation is when the pair $(c,d)$ is {\em{mixed}}: one value is true and the other is false. Then, except for pairs from the same or neighboring blocks, the block order on $U_c\times U_d$ is equal to the adjacency relation or its negation. Therefore, with the help of moduli, we will be able to interpret the block order $\bkleq_{c,d}$ for every mixed pair of essential cells $(c,d)$, and even the block order $\bkleq_c$ for every essential cell $c$ that is involved in any mixed~pair. 

We now proceed to the formal definition of the types.
A pair of (possibly equal) cells $(c,d)\in \Cells\times \Cells$, say $c=(i,X)$ and $d=(j,Y)$, is called:
\begin{itemize}
\item {\em{negative}} if $\phi(j)\notin X$ and $\phi(i)\notin Y$;
\item {\em{positive}} if $\phi(j)\in X$ and $\phi(i)\in Y$; and
\item {\em{mixed}} if $\phi(j)\in X$ and $\phi(i)\notin Y$, or $\phi(j)\notin X$ and $\phi(i)\in Y$.
\end{itemize}
Observe that for a cell $c$, the pair $(c,c)$ is always either positive or negative, never mixed. Let us again stress that the partition of pairs of cells into types depends on the recoloring $\phi$, which is a part of the common abstraction $e$.

The following lemma shows that we can easily define the block order on each mixed pair of essential cells. Recall that a cell $c$ is essential if it belongs to $L$, which means that the set $\sigma_s[c]$ is
nonempty for each $s\in [n]$. Note that for non-essential cells $c$, the set $U_c$ is empty.

\begin{lem}\label{lem:mixed-order}
Suppose $(c,d)$ is a mixed pair of essential cells. Then each of the following relations has interpretation complexity at most $2$
over $\Ga$: $\bkleq_c$, $\bkleq_d$, and $\bkleq_{c,d}$.
\end{lem}
\begin{proof}
As $(c,d)$ is mixed, we have $c\neq d$. Let $c=(i,X)$ and $d=(j,Y)$.
By symmetry, without loss of generality assume that $\phi(j)\notin X$ and $\phi(i)\in Y$.
We first claim that for pairs of vertices from $U_c\times U_d$ that originate in distant blocks, just the adjacency relation defines the block~order.

\begin{clm}\label{cl:distant-cd}
Let $(u,v)\in U_c\times U_d$, and suppose $u\in G_s$ and $v\in G_t$ where $|s-t|>1$. Then $u\bkleq v$ if and only if $u$ and $v$ are adjacent in $G$.
\end{clm}
\begin{proof}
Suppose first that $s<t-1$. Observe that $u$ is colored with $i$ in $\sigma_s$, then it is recolored to $\phi(i)$ when composing with $\sigma_{s+1}$, and it stays colored with $\phi(i)$
when composing with $\sigma_{s+2},\ldots,\sigma_{t-1}$. As $\phi(i)\in Y$ and $v$ resides in cell $(j,Y)$, this means that when composing with $\sigma_t$, we add the edge between $u$ and $v$.

Suppose next that $s>t+1$. Observe that $v$ is colored with $j$ in $\sigma_t$, then it is recolored to $\phi(j)$ when composing with $\sigma_{t+1}$, and it stays colored with $\phi(j)$
when composing with $\sigma_{t+2},\ldots,\sigma_{s-1}$. As
$\phi(j)\notin X$ and $u$ resides in cell $(i,X)$, this means that when composing with $\sigma_s$, we do not add the edge between $u$ and $v$.
\cqed\end{proof}

Next, elements of $U_c$ from distant blocks can be compared using elements of $U_d$.
For $u,v\in U_c$, a vertex $w\in U_d$ is called a {\em{pivot for $(u,v)$}} if the following conditions hold:
\begin{itemize}
\item $\dist(\modulus{u},\modulus{w})>1$, $\dist(\modulus{v},\modulus{w})>1$, and
\item $u$ and $w$ are adjacent, whereas $v$ and $w$ are not adjacent.
\end{itemize}
First, we check that having a pivot implies the block order.

\begin{clm}\label{cl:pivot->leq}
Let $u,v\in U_c$ and suppose there is a pivot for $(u,v)$. Then $u\bkl v$.
\end{clm}
\begin{proof}
Let $w$ be a pivot for $(u,v)$, where $w\in G_p$ for some $p$.
As $\dist(\modulus{u},\modulus{w})>1$ and $\dist(\modulus{v},\modulus{w})>1$, we in particular have that $|s-p|>1$ and $|t-p|>1$.
By Claim~\ref{cl:distant-cd}, the adjacency between $u$ and $w$ implies that $u\bkleq w$, and moreover $u\bkl w$ since $u$ and $w$ have different moduli.
Similarly, the non-adjacency between $v$ and $w$ implies that $w\bkl v$.
By transitivity we infer that $u\bkl v$.
\cqed\end{proof}

Next, we show that vertices of $U_c$ that are in distant blocks always have a pivot.
\begin{clm}\label{cl:leq->pivot}
Let $u,v\in U_c$, and suppose that $u\in G_s$ and $v\in G_t$ where $s<t-3$.
Then there is a pivot for $(u,v)$.
\end{clm}
\begin{proof}
Note that among the $7$ moduli, at most $3$ are equal to or neighboring $\modulus{u}$, and at most $3$ are equal to or neighboring $\modulus{v}$.
Hence there exists a modulus $r$ such that $\dist(\modulus{u},r)>1$ and $\dist(\modulus{v},r)>1$. It is easy to see that since $s<t-3$, there is a number $p$ such that $s+1<p<t-1$ and $p\bmod 7 = r$;
in particular $|s-p|>1$ and $|t-p|>1$.
Since $d$ is an essential cell, there exists a vertex $w$ in the block $G_p$ such that $w\in U_d$. 
By Claim~\ref{cl:distant-cd} we have that $u$ and $w$ are adjacent, while $v$ and $w$ are non-adjacent.
As $\modulus{w}=r$, indeed $w$ is a pivot for $(u,v)$.
\cqed\end{proof}

\begin{figure}[htbp!]
                \centering
		\def\svgwidth{\textwidth}
                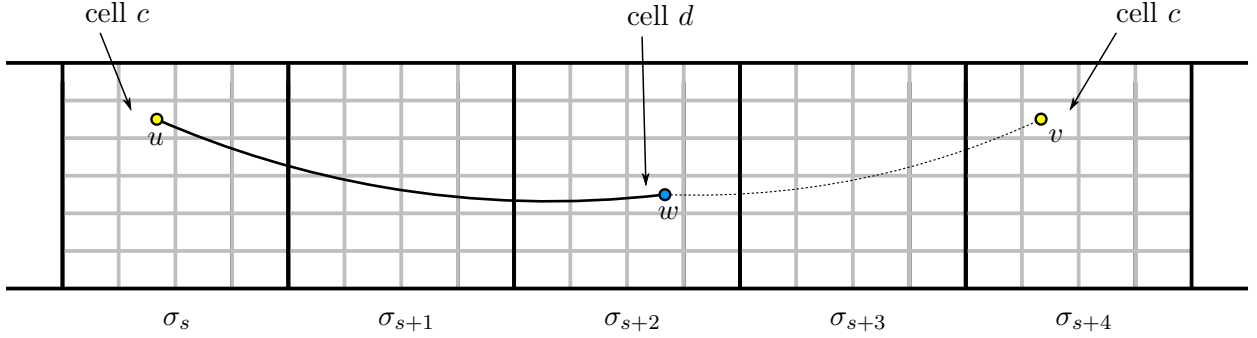
        \caption{A pivot $w$ for vertices $u\in G_s$ and $v\in G_{s+4}$. Solid line between vertices represents an edge, dotted represents a non-edge.}\label{fig:pivot}
\end{figure}

We are ready to interpret $\bkleq_c$.

\begin{clm}
The interpretation complexity of $\bkleq_c$ over $\Ga$ is at most $1$.
\end{clm}
\begin{proof}
Given $u,v\in U_c$, we need to verify whether $u\bkleq v$. 
We first check whether there is a pivot for $(u,v)$ or whether there is a pivot for $(v,u)$. 
If any of these checks holds, then by Claim~\ref{cl:pivot->leq} we can infer whether $u\bkleq v$.
On the other hand, if none of them holds, then by Claim~\ref{cl:leq->pivot} we have that $|s-t|\leq 3$, where $u\in G_s$ and $v\in G_t$.
Then the block order between $u$ and $v$ can be inferred by comparing the moduli of $u$ and $v$: $u\bkleq v$ if and only if 
$\modulus{u}$ is among $\{\modulus{v}-x\colon x\in \{0,1,2,3\}\}$.
It is straightforward to implement this verification by a first-order formula with quantifier rank $1$ working over $\Ga$.
\cqed\end{proof}

A symmetric reasoning yields the following.

\begin{clm}
The interpretation complexity of $\bkleq_d$ over $\Ga$ is at most $1$.
\end{clm}

Finally, we are left with showing that the interpretation complexity of  $\bkleq_{c,d}$ over $\Ga$ is at most $2$.
Given $(u,v)\in U_c\times U_d$, we need to verify whether $u\bkleq v$.
Let $u\in G_s$, $v\in G_t$, $a=\modulus{u}=s\bmod 7$, and $b=\modulus{v}=t\bmod 7$.
Observe first that if $\dist(a,b)>1$, then also $|s-t|>1$, and hence, by Claim~\ref{cl:distant-cd}, the block order between $u$ and $v$ is equivalent to adjacency between $u$ and $v$.
Hence, we are left with considering the case when $\dist(a,b)\leq
1$.

The reasoning will be similar as for interpreting $\bkleq_c$.
Call $w\in V_d$ a {\em{pivot$^\star$}} for the pair $(u,v)$ if $\dist(a,\modulus{w})>1$, $u$ is adjacent to $w$, and $w\bkleq_d v$.
Observe that if there is a pivot$^\star$ $w$ for $(u,v)$, then the adjacency between $u$ and $w$ together with $\dist(a,\modulus{w})>1$ imply, by Claim~\ref{cl:distant-cd}, that $u\bkl w$.
Together with $w\bkleq_d v$, this implies that $u\bkl v$.
Hence, the existence of a pivot$^\star$ for $(u,v)$ implies that $u\bkl v$.

On the other hand, suppose for a moment that $s<t-1$. Then, due to $\dist(a,b)\leq 1$, we actually have $s\leq t-6$.
Hence, there exists some modulus $r$ and index $p$ such that $s+1<p\leq t$, and $r=p\mod 7$, and $\dist(r,a)>1$.
Since cell $d$ is essential, there exists a vertex $w\in G_p$ such that also $w\in U_d$.
Since $s+1<p$, by Claim~\ref{cl:distant-cd} we have that $u$ and $w$ are adjacent, hence $w$ is a pivot$^\star$ for $(u,v)$.

Therefore, in case $\dist(a,b)\leq 1$ we verify whether $u\bkleq v$ as follows. First, check whether there is a pivot$^\star$ for $(u,v)$ or for $(v,u)$.
If any of these checks holds, then this forces the block order between $u$ and $v$, and we can immediately infer whether $u\bkleq v$.
Otherwise there is neither a pivot$^\star$ for $(u,v)$ nor for $(v,u)$, so we conclude that $|s-t|\leq 1$.
Then the block order between $u$ and $v$ can be inferred by comparing the moduli: $u\bkleq v$ if and only if $\modulus{u}$ is equal either to $\modulus{v}$ or to $\modulus{v}-1$.

It is straightforward to implement the above verification using a first-order formula of quantification rank $1$ that uses the formula interpreting $\bkleq_d$.
Since the interpretation complexity of the latter relation is at most $1$, it follows that the interpretation complexity of $\bkleq_{c,d}$ is at most $2$.
\end{proof}

\subsection{Sociability and solitarity}

Lemma~\ref{lem:mixed-order} suggests the following classification of essential cells.
An essential cell $c$ is called {\em{social}} if there is another essential cell $d$ such that $(c,d)$ is a mixed pair of cells.
Essential cells that are not social are called {\em{solitary}}, and a vertex $u$ is {\em{social}} (resp. {\em{solitary}}) if $u\in U_c$ for some social (resp. solitary) cell $c$.
Then Lemma~\ref{lem:mixed-order} asserts that for any social cell $c$, the interpretation complexity of $\bkleq_c$ over $\Ga$ is at most $2$.
Intuitively, our next goal is to extend this order to solitary cells as much as possible, and to piece together the obtained orders $\bkleq_c$ by interpreting the block order between elements from different
cells.

First, we observe that we can in some sense compose the orders $\bkleq_{c,d}$ that we have already interpreted. More precisely,
let $M$ be the {\em{social graph}} defined as follows: the vertex set of $M$ comprises all social cells, and two cells are considered adjacent iff they form a mixed pair.
For a component $C$ of the social graph $M$, we denote $U_C=\bigcup_{c\in C} U_c$.

\begin{lem}\label{lem:paths-social}
Suppose social cells $c$ and $d$ belong to the same connected component of the social graph~$M$. Then the interpretation complexity of $\bkleq_{c,d}$ over $\Ga$ is at most $|\Cells|$.
\end{lem}
\begin{proof}
Let $Q=(c=c_1,c_2,\ldots,c_{p-1},c_p=d)$ be any path in $M$ between $c$ and $d$. Obviously the length of $Q$ is at most $|\Cells|-1$, hence $Q$ has at most $|\Cells|-2$ internal vertices.
Consider the following first-order formula with free variables $u\in U_c$ and $v\in U_d$: there exist vertices $w_2\in V_{c_2}$, $w_3\in V_{c_3}$, and so on up to $w_{p-1}\in V_{c_{p-1}}$, such that
\[u\bkleq_{c,c_2} w_2\bkleq_{c_2,c_3} w_3\bkleq_{c_3,c_4} \ldots \bkleq_{c_{p-3},c_{p-2}} w_{p-2} \bkleq_{c_{p-2},c_{p-1}} w_{p-1}\bkleq_{c_{p-1},d} v.\]
Since all cells $c_2,\ldots,c_{p-1}$ are essential, it is easy to prove
 that $u\bkleq v$ if and only if this formula is satisfied. To see
 this, note that some or all of the $w_i,w_j$ may belong to the same block, which
 guarantees $w_i\bkleq_{c_i,c_j} w_j$.
Moreover, by Lemma~\ref{lem:mixed-order} and the definition of the social graph,
each of the relations $\bkleq_{c_i,c_{i+1}}$ has interpretation complexity at most $2$. Since we quantify at most $|\Cells|-2$ intermediate vertices, it follows that the constructed formula
has quantifier rank at most $|\Cells|$.
\end{proof}

For a connected component $C$ of the social graph $M$, by $\bkleq_C$ we denote the block order restricted to pairs of vertices from $U_C$.
Lemma~\ref{lem:paths-social} immediately yields the following.

\begin{cor}\label{cor:comps-social}
For any connected component $C$ of $M$, the interpretation complexity of $\bkleq_C$ over $\Ga$ is at most $|\Cells|$.
\end{cor}

Finally, for the purpose of further reasoning we need to observe some basic properties of cells.
More precisely, we analyze how the cell of a vertex in its block corresponds to its cell in the overall derivation $\sigma$.
The following assertion follows immediately from the definition of composing derivations and the fact that $\phi$ is an idempotent function.

\begin{obs}\label{obs:overall-cell}
Suppose $u\in G_s$ for some $s\in [n]$ and $u\in U_c$ for some cell $c=(i,X)$.
Then in the derivation $\sigma$, the vertex $u$ belongs to cell $\sigma[c']$ for $c'=(i',X')$ defined as follows:
\[
  i' = \begin{cases}
         i       &\text{if } s = n, \\
         \phi(i) &\text{otherwise;}
       \end{cases}
  \qquad\qquad\text{and}\qquad\qquad
  X' = \begin{cases}
         X            &\text{if } s = 1, \\
         \phi^{-1}(X) &\text{otherwise.}
       \end{cases}
\]
\end{obs}

Next, we verify that cells $c$ and $c'$ as in Observation~\ref{obs:overall-cell} behave in almost the same manner in the social graph $M$.

\begin{lem}\label{lem:cell-projection}
Suppose cells $c=(i,X)$ and $c'=(i',X')$ are such that $i'=i$ or $i'=\phi(i)$, and $X'=X$ or $X'=\phi^{-1}(X)$.
Then for any cell $d$, the pair $(c,d)$ is of the same type---negative, positive, or mixed---as the pair $(c',d)$.
Consequently, $c$ is social if and only if $c'$ is social, and provided $c$ and $c'$ are social, they belong to the same connected component of the social graph~$M$.
\end{lem}
\begin{proof}
Since the second claim of the lemma statement follows immediately from the first claim, we focus on proving the latter. 
Let $d=(j,Y)$.
Observe that $\phi(i')=\phi(i)$, because $\phi$ is idempotent and $i'$ is equal to $i$ or $\phi(i)$.
Observe also that $\phi(j)\in X$ if and only if $\phi(j)\in \phi^{-1}(X)$, again because $\phi$ is idempotent.
Hence, we have $\phi(i)\in Y$ if and only if $\phi(i')\in Y$, and $\phi(j)\in X$ if and only if $\phi(j)\in X'$, because $X'$ is equal either to $X$ or to $\phi^{-1}(X)$.
It follows that the pairs $(c,d)$ and $(c',d)$ are of the same type.
\end{proof}

Note that Lemma~\ref{lem:cell-projection} in particular applies to $d=c$ and $d=c'$. 
We can thus infer that pairs $(c,c)$, $(c,c')$, and $(c',c')$ are always of the same type.

\subsection{Flipping the graph}

We now came to the point where we can define the set $Z$ whose existence is postulated in the conclusion of the Definable Order Lemma. Recall that our goal is to give an upper bound on the interpretation complexity of the block order restricted to pairs belonging to the same connected component of the $Z$-flip of $G$.
The idea is very simple: we take $Z\subseteq \binom{\Cells}{1,2}$ to be the set of all positive pairs of cells.  That is, for each pair $(c,d)$ of essential cells (possibly $c=d$), if $(c,d)$ is positive then we put $\{c,d\}$ into $Z$. The intuition behind this choice is as follows. Essentially, one can imagine that after the flip, all pairs of essential cells will be either mixed or negative, because positive pairs get flipped into negative ones. For mixed pairs of cells we have already defined the block order. However, as argued in the beginning of Section~\ref{sec:types}, if $(u,v)\in U_c\times U_d$ where $(c,d)$ is a negative pair, then $u$ and $v$ may be adjacent only if $u$ and $v$ belong to the same or neighboring blocks. Thus, after the flip all adjacencies are either ``already understood'' (for mixed pairs) or ``local'' (for negative pairs). Very roughly speaking, the locality of the latter adjacencies allows us to understand them by making connectivity queries to the $Z$-flip of $G$.

We now proceed with formalizing this intuition.
Having defined $Z$ as the set of all positive pairs of cells,
we construct a graph $H$ from $G$ by performing the flip between $U_c$ and $U_d$ for every $\{c,d\}\in Z$, and performing the flip between $U_c$ and $U_c$ for every $\{c\}\in Z$. Note that a priori it is not clear that $H$ is the $Z$-flip of $G$, as $U_c$ and $U_d$ are not necessarily equal to $\sigma[c]$ and $\sigma[d]$. However, from Lemma~\ref{lem:cell-projection} one can easily infer that this is the case; this is checked formally in the following lemma.

\begin{lem}\label{lem:flip-same}
The $Z$-flip of $\sigma$ is equal to $H$.
\end{lem}
\begin{proof}
Let $H'$ be the $Z$-flip of $\sigma$.
It suffices to show that a pair of vertices $u,v$ is adjacent in $H$ if and only if it is adjacent in $H'$.
Let $u\in U_c$, $u\in \sigma[c']$, $v\in U_d$, and $v\in \sigma[d']$; here, the pairs of cells $(c,c')$ and $(d,d')$ are as described in Observation~\ref{obs:overall-cell}.
By applying Lemma~\ref{lem:cell-projection} twice, we have that $(c,d)$ is of the same type as $(c',d)$, which in turn is of the same type as $(c',d')$.
If $(c,d)$ is negative, then $(c',d')$ is also negative, and the adjacency between $u$ and $v$ from $G$ is negated neither when constructing $H$ nor when constructing $H'$.
Consequently, $u$ and $v$ are adjacent in $H$ iff they are adjacent in $G$ iff they are adjacent in $H'$.
The same holds also when $(c,d)$ (equivalently, $(c',d')$) is mixed.
However, when $(c,d)$ is positive, then $(c',d')$ is also positive, so the adjacency between $u$ and $v$ is negated both when constructing $H$ and when constructing $H'$.
Consequently, $u$ and $v$ are adjacent in $H$ iff they are non-adjacent in $G$ iff they are adjacent in $H'$.
\end{proof}

The next lemma shows that the performed flipping operation indeed simplifies the adjacency relation in the graph, as explained in the intuition at the beginning of this subsection.

\begin{lem}\label{lem:neighboring-blocks}
Suppose $u$ and $v$ are vertices adjacent in $H$, where $u\in G_s\cap U_c$ and $v\in G_t\cap U_d$ for some $s,t\in [n]$ and $c,d\in \Cells$; possibly $c=d$.
Suppose further that $(c,d)$ is not a mixed pair.
Then $|s-t|\leq 1$.
\end{lem}
\begin{proof}
For the sake of contradiction, suppose that $|s-t|>1$. By symmetry, we
may further assume $s<t-1$.
Let $c=(i,X)$ and $d=(j,Y)$.

Observe that vertex $u$ was colored with color $i$ in $\sigma_s$, then it was recolored to $\phi(i)$ when composing with $\sigma_{s+1}$, and its color stayed equal to $\phi(i)$
when composing with $\sigma_{s+2},\ldots,\sigma_{t-1}$. Hence, when composing with $\sigma_t$, the edge between $u$ and $v$ was added in $G$ if and only if $\phi(i)\in Y$.
As $(c,d)$ is not a mixed pair, it is either positive or negative.
If $(c,d)$ is negative, then $\phi(i)\notin Y$, $u$ and $v$ were not adjacent in $G$ and there was no flipping between $U_c$ and $U_d$ when constructing $H$.
Consequently $u$ and $v$ stay non-adjacent in $H$; a contradiction.
Otherwise, if $(c,d)$ is positive, then $\phi(i)\in Y$, $u$ and $v$ are adjacent in $G$, but we applied flipping between $U_c$ and $U_d$ when constructing $H$.
Consequently $u$ and $v$ are non-adjacent in~$H$; a contradiction again.
\end{proof}

Lemma~\ref{lem:neighboring-blocks} tells us that apart from adjacencies in mixed pairs, the adjacency relation in~$H$ is local: vertices adjacent in $H$ lie either in the same or in neighboring blocks,
provided the relation between their cells is not mixed.
Note that this prerequisite is satisfied always when at least one of the vertices is solitary. 

In the remainder of the proof we will predominantly work with the graph $H$, hence for the purpose of interpreting various relations, it will be useful to have it encoded in the considered relational structure.
Let then $\Ha$ be the structure obtained from $\Ga$ by expanding it by the adjacency relation of $H$; note that $\Ha$ still contains the adjacency relation of $G$ as well as unary predicates for sets $U_c$ and moduli.
Observe that the adjacency relation in $H$ can be interpreted by a first-order formula of quantifier rank $0$ working on $\Ga$, so its interpretation complexity over $\Ga$ is $0$.
Hence, from now on we may assume that we work over $\Ha$.

\subsection{Reduction to a single connected component}

Recall that our ultimate goal is to give an upper bound on the interpretation complexity of $\sim\cap \bkleq$, where $\sim$ is the equivalence relation of being in the same connected component of $H$. However, it would be more convenient to focus on a single connected component of $H$, that is, on interpreting the block order within this component. The goal of this subsection is to formally prove that we can restrict attention to this case. 

For a connected component $F$ of $H$, by $\bkleq_F$ we denote the block order restricted to the vertices of $F$. In the remainder of the proof, we will argue the following statement.

\begin{lem}\label{lem:comps-of-H}
For each connected component $F$ of $H$, the interpretation complexity of $\bkleq_F$
over $\Ha$ expanded by a unary predicate selecting the vertices of $F$ is at most $3|\Cells|+5$.
\end{lem}

While Lemma~\ref{lem:comps-of-H} will be proved in the subsequent subsections, we now show that it implies the Definable Order Lemma.

\begin{proof}[Proof of the Definable Order Lemma assuming Lemma~\ref{lem:comps-of-H}]
Recall that by Lemma~\ref{lem:flip-same} we know that $H$ is equal to the $Z$-flip of $\sigma$.
As argued, we may assume that we work over $\Ha$ instead of the original~$G$.

We now need to write an \mso formula that for given vertices $u,v$ checks whether $u$ and $v$ are in the same connected component of $H$ and moreover $u\bkleq v$.
The assertion that $u$ and $v$ are in the same connected component of $H$ is clearly expressible in \mso, as the adjacency relation of $H$ is present in the structure.
For the assertion $u\bkleq v$ we will use an appropriate interpretation provided by Lemma~\ref{lem:comps-of-H}.
More precisely, suppose $F$ is the connected component of $H$ that contains $u$ and $v$.
Then Lemma~\ref{lem:comps-of-H} asserts that the interpretation complexity of $\bkleq$ restricted to $F$ is at most $3|\Cells|+5$; more precisely, there exists a formula $\varphi_F$
of quantifier rank at most $3|\Cells|+5$ and using at most $3|\Cells|+5$ monadic variables that interprets $\bkleq$ restricted to $F$.
Note that the number of possible such formulas $\varphi_F$ is bounded by a function of $k$.
Hence, for every such possible formula $\psi$ we may introduce an (existentially guessed) monadic parameter $X_\psi$ that selects the union of vertex sets of those connected components $F$ of $H$ for which
$\varphi_F=\psi$. Then to check whether $u\bkleq v$ one may use the formula $\psi$ for which $u,v\in X_\psi$~holds.
\end{proof}

Thus, from now on we may focus on proving Lemma~\ref{lem:comps-of-H}.

\subsection{Solitary paths}

The proof of Lemma~\ref{lem:comps-of-H} is divided into several steps.
Recall that, as shown in Corollary~\ref{cor:comps-social}, for each connected component $C$ of the social graph $M$ we have already interpreted the block order within cells that belong to $C$. The intuition is that now, we try to ``connect'' different components of $M$ with each other, and to ``connect'' solitary vertices (i.e. vertices belonging to solitary cells) to social vertices (i.e. vertices belonging to social cells, and thus participating in the connected components of $M$). 

The key tool in formalizing these ``connections'' will be the notion of a {\em{solitary path}}, which is essentially a path in $H$ all of whose  internal vertices are solitary. One the one hand, solitary paths are basic ``links'' that can connect two social vertices, or a solitary vertex with a social vertex. On the other hand, by Lemma~\ref{lem:neighboring-blocks} we know that such paths cannot jump between distant blocks, hence we will be able to control guessing them in \mso with the help of moduli.
We will also use the idempotence of the abstraction $e=\abst{\sigma}$, which is equal to $\abst{\sigma_s}$ for all $s\in [n]$, to reason about the existence of some ``local'' solitary paths realizing sought connections.

Our goal in this subsection is to work out basic properties of solitary paths. Formally, a path $Q$ in $H$ is called {\em{solitary}} if the following conditions~hold:
\begin{itemize}
\item all internal vertices of $Q$ are solitary; and 
\item if the endpoints of $Q$ belong to $U_c$ and $U_d$, respectively, then $(c,d)$ is not a mixed pair.
\end{itemize}
Note that the endpoints of a solitary path may be social, but we explicitly exclude the case when they belong to cells forming a mixed pair.
The following assertion about locality of solitary paths follows immediately from Lemma~\ref{lem:neighboring-blocks}.

\begin{obs}\label{obs:local-solitary}
If $Q$ is a solitary path, then any two consecutive vertices on $Q$ belong either to the same or to neighboring blocks.
\end{obs}

We first show that when two different components of the social graph $M$ can be connected by a solitary path, then they can be connected by a solitary path that is entirely contained in one block, and even this block can be chosen arbitrarily.
For this we crucially use the idempotence of abstraction $e$.

\begin{lem}\label{lem:local-connection}
Suppose $C,D$ are two different connected components of the social graph $M$.
Suppose further that there is a solitary path in $H$ that starts in a vertex of $U_C$ and ends in a vertex of $U_D$.
Then, for each $s\in [n]$, there is a solitary path that starts in an vertex of $U_C$, ends in a vertex of $U_D$, and all of whose vertices belong to the block $G_s$.
\end{lem}
\begin{proof}
Let $Q$ be the path whose existence is asserted in the statement of the lemma.
Consider $Q$ as a path in the $Z$-flip of $\sigma$, which is equal to $H$ by Lemma~\ref{lem:flip-same}.
Suppose $Q$ starts in $\sigma[c']$, ends in $\sigma[d']$, and the set of cells of $\sigma$ traversed by the internal vertices of $Q$ is $W'$.
Since $Q$ is solitary, by Lemma~\ref{lem:cell-projection} we infer that all the cells of $W'$ are solitary.
The endpoints of $Q$ belong then to $U_c$ and $U_d$ for some $c,d\in \Cells$, respectively, such that $c\in C$, $d\in D$, and the pairs $c,c'$ and $d,d'$ are as described in Observation~\ref{obs:overall-cell}.
In particular, by Lemma~\ref{lem:cell-projection} we have that $c'\in C$ and~$d'\in D$.

Since $\abst{\sigma}=\abst{\sigma_s}$ for each $s\in [n]$, the connectivity registries of these abstractions are also equal; recall from Definition~\ref{def:derivation} that the connectivity registry is the component of an abstraction that holds information about the connectivity between cells in flips of the underlying graph.
Hence, 
the existence of $Q$ implies the existence of a path $Q'$ in the $Z$-flip of $\sigma_s$ such that $Q'$ starts in $\sigma_s[c']$, ends in $\sigma_s[d']$, 
and all the internal vertices of $Q'$ belong to cells $\sigma_s[b]$ for $b\in W'$.
This means that $Q'$ starts in $U_{c'}\subseteq U_C$, ends in $U_{d'}\subseteq U_D$, and each of its internal vertices belongs to some $U_b$ for $b\in W'$, which implies that it is solitary.
Since $(c,d)$ is not mixed due to $Q$ being solitary, we infer that $(c',d')$ is also not mixed by Lemma~\ref{lem:cell-projection}.
Hence $Q'$ is a solitary path and we are done.
\end{proof}

Finally, we show that if a solitary vertex $u$ can be connected to a component of the social graph $M$ via a solitary path, 
then there is also such a solitary path that that may traverse only the block to which $u$ belongs, the preceding block, and the succeeding block. On the intuitive level, this, together Lemma~\ref{lem:local-connection}, shows that connections via solitary paths can be made local in the sense that the path traverses only a constant number of consecutive blocks.

\begin{lem}\label{lem:sol-to-social}
Suppose $u\in G_s$ is a solitary vertex and there is a solitary path $Q$ in $H$ that leads from $u$ to some social vertex $v$, say belonging to $U_D$ for some connected component $D$ of the social graph $M$.
Then there is also a solitary path $Q'$ in $H$ that leads from $u$ to some social vertex $v'$ belonging to $U_D$ such that all vertices traversed by $Q'$ belong to the blocks $G_{s-1}$, $G_s$, and $G_{s+1}$.
\end{lem}
\begin{proof}
By changing $Q$ if necessary, we may assume that out of solitary paths that lead from $u$ to any social vertex from $U_D$, $Q$ is a path that uses the smallest number of vertices outside of the blocks
$G_{s-1}$, $G_s$, and $G_{s+1}$. It suffices to show that $Q$ chosen in this manner in fact traverses only vertices from these blocks, so assume otherwise.
Let $d\in D$ be such that~$v\in U_d$.

In the following, we regard $Q$ as traversed in the direction from $u$ to $v$.
Let $r$ be any vertex on $Q$ that does not belong to any of the blocks $G_{s-1}$, $G_s$, and $G_{s+1}$.
Let $x$ be the earliest vertex before $r$ on $Q$ with the following property: all the vertices between $x$ and $r$ (inclusive) on $Q$ do not belong to $G_s$.
Similarly, let $y$ be the latest vertex after $r$ on $Q$ with the following property: all the vertices between $r$ and $y$ (inclusive) on $Q$ do not belong to $G_s$.
Note that $x$ has a predecessor on $Q$ that belongs to $G_s$, while $y$ either has a successor on $Q$ that belongs to $G_s$, or $y=v$.
Let $R$ be the infix of $Q$ between $x$ and $y$ (inclusive).
Note that~$x$ is an internal vertex of $Q$, hence it is solitary.

By Lemma~\ref{lem:neighboring-blocks}, we know that every pair of consecutive vertices on $Q$ either belong to the same or to neighboring blocks.
Consequently, by the definition of $R$ we have that one of the following two cases holds:
\begin{enumerate}[(1)]
\item\label{alt:-} all vertices on $R$ belong to $\bigcup_{t<s} G_t$, $x\in G_{s-1}$, and (either $y=v$ or $y\in G_{s-1}$); or
\item\label{alt:+} all vertices on $R$ belong to $\bigcup_{t>s} G_t$, $x\in G_{s+1}$, and (either $y=v$ or $y\in G_{s+1}$).
\end{enumerate}
To prove the lemma, it suffices to show the following claim about the existence of a suitable replacement path for the infix $R$.
See Figure~\ref{fig:replacement} for reference.

\begin{figure}[htbp!]
                \centering
		\def\svgwidth{0.8\textwidth}
                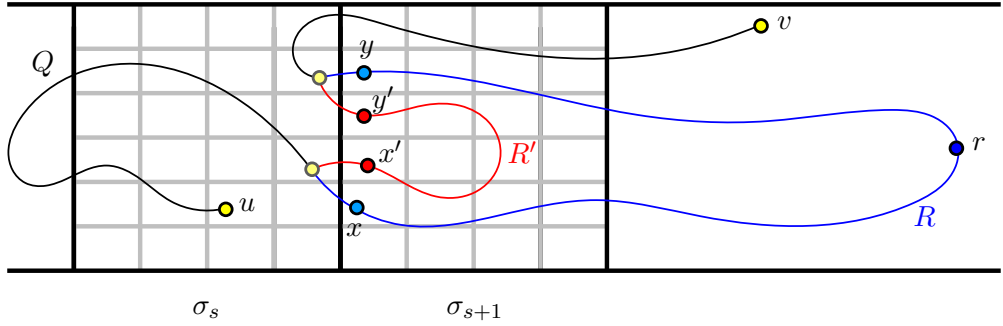
        \caption{The situation in Claim~\ref{cl:replacement} for the alternative~\ref{alt:+}. The replaced path $R$ is depicted in blue, the replacement path $R'$ is depicted in red.}\label{fig:replacement}
\end{figure}

\begin{clm}\label{cl:replacement}
There exists a solitary path $R'$, say with endpoints $x'$ and $y'$, that satisfies the following:
\begin{enumerate}[(a)]
\item\label{p:same-block} all vertices traversed by $R'$ belong to the same block as $x$;
\item\label{p:x} $x'$ is solitary and is adjacent in $H$ to the predecessor of $x$ on $Q$;
\item\label{p:yv} if $y=v$, then $y'$ belongs to $U_D$;
\item\label{p:v} if $y\neq v$, then $y'$ is solitary and is adjacent in $H$ to the successor of $y$ on $Q$.
\end{enumerate}
\end{clm}

Observe that it suffices to prove Claim~\ref{cl:replacement} for the following reason.
Take $Q'$ to be $Q$ with the infix $R$ replaced with $R'$.
By~\ref{p:x},~\ref{p:yv}, and~\ref{p:v}, $Q'$ constructed in this manner is still a solitary path in $H$, and it ends in a vertex of $U_D$.
However, by~\ref{p:same-block}, $Q'$ traverses strictly fewer vertices outside of blocks $G_{s-1}$, $G_s$, and $G_{s+1}$, because $R'$ is entirely contained in these blocks, while $R$ traversed
vertex $r$ which lies outside of these blocks. Thus, the existence of $Q'$ contradicts the initial choice of $Q$.

Hence, from now on we focus on proving Claim~\ref{cl:replacement}. We consider two cases: either alternative~\ref{alt:-} holds, or alternative~\ref{alt:+} holds.

Suppose first that alternative~\ref{alt:+} holds: $R$ is contained in $\bigcup_{t>s} G_t$, $x\in G_{s+1}$, and (either $y=v$ or $y\in G_{s+1}$).
Consider the derivation
\[\sigma'=\sigma_{s+1}\cdot \sigma_{s+2}\cdot \ldots\cdot \sigma_n.\]
Since the abstraction $e$ is idempotent, we have that $\abst{\sigma'}=e$, in particular $\abst{\sigma'}$ has the same connectivity registry as $\abst{\sigma_t}$ for each $t\in [n]$.
Let $c_x,c_y,c_x',c_y'\in \Cells$ be such that $x\in U_{c_x}$, $x\in \sigma'[c_x']$, $y\in U_{c_y}$, and $y\in \sigma'[c_y']$.
By Observation~\ref{obs:overall-cell}, the cells $c_x$ and $c_x'$ have the same profile, say $X$, but may differ in the color.
Similarly, the cells $c_y$ and $c_{y'}$ have the same profile, say $Y$, but may differ in the~color.

Consider now the path $R$ as a path in the $Z$-flip of the derivation $\sigma'$, which is equal to the subgraph induced in $H$ by the vertex set of the underlying graph of $\sigma'$, by Lemma~\ref{lem:flip-same}.
This path starts in $\sigma'[c_x']$, ends in $\sigma'[c_y']$, 
and let $W$ be the set of cells in $\sigma'$ traversed by the internal vertices of $R$. 
Note that since $R$ is solitary, all the internal vertices of $R$ are solitary, hence by Lemma~\ref{lem:cell-projection} we have that all the cells of $W$ are solitary.
Observe now that the existence of $R$ and the fact that the connectivity registries of $\sigma'$ and $\sigma_{s+1}$ are equal certify that there is a path $R'$ in the $Z$-flip of $\sigma_{s+1}$
that starts in $\sigma_{s+1}[c_x']$, ends in $\sigma_{s+1}[c_y']$, and travels through cells of $W$. We claim that $R'$ satisfies all the required properties.

First, observe that $R'$ is a path in the $Z$-flip of $\sigma_{s+1}$, which is equal to the subgraph induced in $H$ by the vertex set of $G_{s+1}$.
Thus, $R'$ is a path in $H$ with all vertices belonging to $G_{s+1}$.
It starts with some vertex $x'$ belonging to $U_{c_x'}$.
Note here that, by Lemma~\ref{lem:cell-projection}, $c_x'$ is solitary because $c_x$ is solitary, hence $x'$ is solitary as claimed.
Similarly, $R'$ ends in some vertex $y'$ belonging to $U_{c_y'}$. Note that $y'$ is solitary if and only if $y$ is solitary, which happened if and only if $y\neq v$.
Observe further that $R'$ is solitary because all the internal vertices traversed by $R'$ belong to sets $\sigma_{s+1}[b]\subseteq U_b$ for $b\in W$, and all the cells of $W$ are solitary.

Since the profiles of $x$ and $x'$ are equal, the predecessor of $x$ on $Q$ is adjacent to $x'$ in $G$ if and only if it is adjacent to $x$ in $G$.
Supposing this predecessor belongs to $U_b$ for some $b\in \Cells$, by Lemma~\ref{lem:cell-projection} we infer that type of the pair $(b,c_x)$ is the same as the type of $(b,c_x')$.
Consequently, the adjacency between the predecessor and $x$ is negated when constructing $H$ if and only if the adjacency between the predecessor and $x'$ is negated.
As the predecessor and $x$ are adjacent in $H$, we infer that the predecessor is adjacent also to $x'$ in $H$.
A symmetric reasoning shows that provided $y\neq v$, the successor of $y$ on $Q$ is adjacent to $y'$ in $H$.
Finally, observe that if $y=v$, then since $y'\in U_{c_y'}$ and $c_y,c_y'$ belong to the same connected component of the social graph $M$ (by Lemma~\ref{lem:cell-projection}), 
we have that $y'\in U_D$. This concludes the verification that $R'$ has all the claimed properties.

Next, consider the alternative~\ref{alt:-}: $R$ is contained in $\bigcup_{t<s} G_t$, $x\in G_{s-1}$, and (either $y=v$ or $y\in G_{s-1}$).
The proof is almost entirely symmetric; we simply consider the derivation $\sigma'=\sigma_1\cdot \ldots\cdot \sigma_{s-1}$, and use the equality of connectivity registries of $\abst{\sigma'}$ and $\abst{\sigma_{s-1}}$.
The only different detail is the verification that $x'$ is adjacent to the predecessor of $x$ on $Q$, and that $y'$ is adjacent to the successor of $y$ on $Q$ (provided $y\neq v$).
Now, by Observation~\ref{obs:overall-cell} we have that $x$ has the same color in $\sigma'$ and in $\sigma_{s-1}$, not the same profile as before.
Consequently, the path $R'$ is chosen so that $x$ and $x'$ have the same color in $\sigma'$, which implies that the predecessor of $x$ on $Q$ is adjacent to $x'$ in $G$ if and only if it is adjacent to $x$ in $G$,
because this assertion is equivalent to the common color of $x$ and $x'$ belonging to the profile of the predecessor.
The rest of the reasoning, including the $y$-counterpart, is exactly symmetric; we leave the verification to the~reader.
\end{proof}

\subsection{Clusters}

With solitary paths understood, we can further extend our interpretation of the block order. In this subsection we will do it for {\em{clusters}}, which are sets of the connected components of the social graph that can be linked using solitary paths.

We say that two connected components $C$ and $D$ of the social graph $M$ are {\em{close}} if there is a solitary path in $H$ that starts in a vertex of $U_C$ and ends in a vertex of $U_D$.
Consider a graph with the connected components of $M$ as the vertex set, where two components are considered adjacent whenever they are close.
The connected components of this graph will be called {\em{clusters}}, and we identify each cluster $A$ with the union of the sets $U_C$ for $C$ belonging to~$A$. 
For a cluster $A$, by $\bkleq_A$ we denote the restriction of the block order $\bkleq$ to the vertices of~$A$.

\begin{lem}\label{lem:cluster-order}
For a cluster $A$, the interpretation complexity of $\bkleq_A$ over $\Ha$ is at most $3|\Cells|$.
\end{lem}
\begin{proof}
Consider an \mso formula with free vertex variables $u$ and $v$ expressing the following property: 
For some $p\leq |\Cells|$, there exist components $\{C^i\colon i\in [p]\}$ of $M$, vertices $\{x_i,y_i\colon i\in [p-1]\}$, and paths $\{Q_i\colon i\in [p-1]\}$ in $H$ such that the following conditions are satisfied:
\begin{itemize}
\item for each $i\in [p-1]$, we have that $x_i\in U_{C^i}$ and $y_i\in U_{C^{i+1}}$;
\item denoting $y_0=u$ and $x_p=v$, for each $i\in [p]$ we have $y_{i-1}\bkleq_{C^i} x_i$; and
\item each path $Q_i$ is solitary, has endpoints $x_i$ and $y_i$, and all vertices traversed by it have the same modulus.
\end{itemize}
Since the social graph $M$ is fixed, and the interpretation complexity of each relation $\bkleq_{C^i}$ is at most $|\Cells|$ by Corollary~\ref{cor:comps-social},
it is straightforward to construct such a formula of quantifier rank at most $3|\Cells|$. We now verify that this formula holds for a pair of vertices $u,v\in A$ if and only if $u\bkleq v$, which will conclude the proof.

First, suppose that indeed $u\bkleq v$; say $u\in G_s$ and $v\in G_t$, where $s\leq t$. Since $u$ and~$v$ both belong to the cluster $A$, there is some $p\leq |\Cells|$ and components $C^1,C^2,\ldots,C^p$ of $M$ such that $u\in U_{C^1}$, $v\in U_{C^p}$, and $C^i$ is close to $C^{i+1}$ for each $i\in [p-1]$.
Let $Q_i$ be a path that certifies this closeness; that is, $Q_i$ is a solitary path that starts in $U_{C^i}$ and ends in $U_{C^{i+1}}$.
By Lemma~\ref{lem:local-connection}, we can choose the paths $Q_i$ so that all their vertices are contained in $G_s$ (and thus have the same modulus). In particular, if we name the endpoints of $Q_i$ as $x_i$ and $y_i$, where $x_i\in U_{C^i}$ and $y_i\in U_{C^{i+1}}$, then all the vertices $\{x_i,y_i\colon i\in [p-1]\}$ belong to $G_s$. Therefore, for all $i\in [p-1]$ we in fact have $y_{i-1}\bkeq x_i$ (implying $y_{i-1}\bkleq_{C^i} x_i$), while that $y_{p-1}\bkleq_{C^p} x_p=v$ follows from $s\leq t$. Thus, components $\{C^i\colon i\in [p]\}$ of $M$, vertices $\{x_i,y_i\colon i\in [p-1]\}$, and paths $\{Q_i\colon i\in [p-1]\}$ together witness that the formula holds for $u$ and $v$.

Second, we need to prove that if the formula holds for some $u,v\in A$, then we have~$u\bkleq v$.
To this end, fix components $\{C^i\colon i\in [p]\}$, vertices $\{x_i,y_i\colon i\in [p-1]\}$, and paths $\{Q_i\colon i\in [p-1]\}$ that 
witness the satisfaction of the formula for $u$ and $v$.
Each $Q_i$ is a solitary path whose vertices have all the same modulus. 
Hence, from Observation~\ref{obs:local-solitary} it follows that $Q_i$ must be entirely contained in one block, hence $x_i\bkeq y_i$ for all $i\in [p-1]$.
We conclude~that
\[u\bkleq_{C^1} x_1\bkeq y_1\bkleq_{C^2} x_2\bkeq y_2 \bkleq_{C^3}\ldots \bkleq_{C^{p-2}} x_{p-1} \bkeq y_{p-1} \bkleq_{C^{p-1}} v,\]
so indeed $u\bkleq v$.
\end{proof}

Before we proceed to the proof of Lemma~\ref{lem:comps-of-H}, we need one more claim about the relation between clusters and connected components of $H$.

\begin{lem}\label{lem:one-cluster}
For each connected component $F$ of $H$, there is at most one cluster
that has a nonempty intersection with $F$.
\end{lem}
\begin{proof}
It suffices to prove that any two social vertices $u,v\in F$ belong to the same cluster.
Let $Q$ be a path in $H$ with endpoints $u$ and $v$, and let $u=w_1,w_2,\ldots,w_p=v$ be the social vertices traversed by $Q$, in the order of traversal.
For each $i\in [p]$, let $C^i$ be the component of the social graph $M$ such that $w_i\in U_{C^i}$, and let $w_i\in U_{c^i}$ for some $c^i\in C^i$.
If $(c^i,c^{i+1})$ is a mixed pair, then $c^i$ and $c^{i+1}$ are adjacent in $M$, and consequently $C^i=C^{i+1}$.
Otherwise, the infix of $Q$ between $w_i$ and $w_{i+1}$ is a solitary path that witnesses that $C^i$ and $C^{i+1}$ are close.
In conclusion, for each $i\in [p-1]$ we have shown that $C^i$ and $C^{i+1}$ are either close or equal, hence all the components $C^i$ for $i\in [p]$ belong to the same cluster.
\end{proof}

\subsection{Proving Lemma~\ref{lem:comps-of-H}}

We can finally piece together all the prepared tools and give a proof of Lemma~\ref{lem:comps-of-H}. We separate two cases: either the considered component $F$ of $H$ contains some social vertex, or it consists only of solitary vertices. In the first case, we will use the interpretation given by Lemma~\ref{lem:cluster-order} together with the local connections given by Lemma~\ref{lem:sol-to-social}. In the second case, we give a more direct argument, which again relies on the locality of adjacencies between solitary vertices.

\begin{lem}\label{lem:comps-of-H-soc}
Suppose $F$ is a connected component of $H$ that contains some social vertex. Then the interpretation complexity of $\bkleq_F$ over $\Ha$ expanded by a unary predicate selecting vertices of $F$
is at most $3|\Cells|+5$.
\end{lem}
\begin{proof}
Since $F$ contains some social vertex, there is some cluster $A$ that intersects $F$. By Lemma~\ref{lem:one-cluster}, this cluster $A$ is unique.
Observe that one can check using an \mso of quantifier rank~$4$ that a given subset of cells forms this unique cluster $A$. Hence, by making a disjunction over all subsets of cells, we can assume further that the constructed formula
may use an additional unary predicate that selects the vertices of $A$.

We would like to construct an \mso formula that for given $u,v\in F$, verifies whether~$u\bkleq v$.
For simplicity of presentation, in the construction we assume that $u$ and $v$ are solitary. The construction of the formula in the other cases is even simpler and we discuss it at the~end.

As argued, the constructed formula is allowed to work over the structure $\Ha$ expanded by two unary predicates that select the vertices of $A$ and $F$, respectively. Given vertices $u$ and $v$ as free variables, the formula expresses the following properties.
\begin{itemize}
\item There exists a vertex $x\in A$ that can be connected with $u$ by a solitary path traversing only vertices of the same or neighboring moduli as $u$.
\item There exists a vertex $y\in A$ that can be connected with $v$ by a solitary path traversing only vertices of the same or neighboring moduli as $v$.
\item There exists a vertex $x'\in A$ such that $x$ and $x'$ belong to
  the same block if the moduli of $u$ and $x$ are equal, $x'$ belongs to the block immediately before the block of $x$ in case the modulus of $x$
is one larger than the modulus of $u$, and $x'$ belongs to the block immediately after the block of $x$ in case the modulus of $x$ is one smaller than the modulus of $u$.
\item There exists a vertex $y'\in A$ such that $y$ and $y'$ belong to
  the same block if the moduli of $u$ and $y$ are equal, $y'$ belongs to the block immediately before the block of $y$ in case the modulus of $y$
is one larger than the modulus of $v$, and $y'$ belongs to the block immediately after the block of $y$ in case the modulus of $y$ is one smaller than the modulus of $v$.
\item It holds that $x'\bkleq_A y'$.
\end{itemize}
Having existentially quantified $x,y$, the first two properties can be easily checked using formulas of quantifier rank $2$.
Then, having existentially quantified $x',y'$, the next two properties can be checked using formulas of quantifier rank $1$ that use the relation $\bkleq_A$, so $1+3|\Cells|$ in total by Lemma~\ref{lem:cluster-order}.
Indeed, for example to check whether $x'$ is in the block immediately after the block $x$, it suffices to make sure that $x\bkl_A x'$ and there is no vertex $x''\in A$ such that $x\bkl_A x''\bkl_A x'$.
The last check requires quantifier rank $3|\Cells|$, by Lemma~\ref{lem:cluster-order}.

Thus, the quantifier rank of the formula is at most $3|\Cells|+5$, as required. We now  verify that the formula holds for solitary vertices $u,v\in F$ if and only if $u\bkleq v$.

First, suppose that $u\bkleq v$; say $u\in G_s$ and $v\in G_t$, where $s\leq t$. Observe that there are solitary paths $Q$ and $R$ in $H$ such that $Q$ connects $u$ with some social vertex $x\in A$, and $R$ connects $v$ with some social vertex $y\in A$.
Indeed, it suffices to take shortest paths from $u$ and $v$ to the set of social vertices in $H$.
By Lemma~\ref{lem:sol-to-social}, we can choose $Q$ so that it traverses only vertices from $G_{s-1}$, $G_s$, and $G_{s+1}$; in particular $x$ belongs to one of these blocks.
Similarly, we can choose $R$ so that it traverses only vertices from $G_{t-1}$, $G_t$, and $G_{t+1}$; in particular $y$ belongs to one of these blocks. Let $x'$ be any vertex that belongs $\sigma_s[c]$, where $c$ is the cell such that $x\in U_c$. Note that the existence of $x$ witnesses that $c$ is essential, hence $\sigma_s[c]$ is nonempty. Similarly, we define $y'$ to be any vertex of $\sigma_t[d]$, where $d$ is the cell such that $y\in U_d$. Observe that since $x,x'\in U_c$, $y,y'\in U_d$, and $x,y\in A$, we also have $x',y'\in A$. Further, as $x'\in G_s$, $y'\in G_t$, and $s\leq t$, we have $x'\bkleq_A y'$. Thus, vertices $x,x',y,y'$ witness that the formula is satisfied for $u$ and $v$.

Second, we argue that if the formula is satisfied, then $u\bkleq v$.
Let $x,x',y,y'$ be vertices whose existence is asserted by the satisfaction of the formula, and let $Q,R$ be the solitary paths witnessing the satisfaction of the first two properties.
Since $Q$ is solitary, by Observation~\ref{obs:local-solitary}, every two consecutive vertices on $Q$ belong to the same or to neighboring blocks. Since the vertices of $Q$ have only one of three moduli:
the modulus of $u$ or the two neighboring ones, it follows that all vertices of $Q$ in fact must be contained in the blocks $G_{s-1}$, $G_s$, or~$G_{s+1}$. Then the quantification of $x'$ ensures
that $x'\in G_s$. Indeed, if for instance $x\in G_{s-1}$, then the modulus of $x$ is one smaller than the modulus of~$u$, hence $x'$ is chosen from the block immediately after the block of $x$, that is, from $G_s$. The other two cases --- $x\in G_s$ and $x\in G_{s+1}$ --- are analogous. A symmetric reasoning shows that $y'\in G_t$. Hence, $u\bkleq v$ is equivalent to $x'\bkleq_A y'$, which we check in the last property.

To conclude the proof, it remains to argue what happens if one or two of the vertices $u,v$ are social. In case both of them are social, we can simply use the already interpreted relation $\bkleq_A$.
In case one of them, say $u$, is social, we proceed exactly as above, except we put $u=x=x'$ instead of finding $x,x'$ via an existential guess of the path $Q$.
\end{proof}

\begin{lem}\label{lem:comps-of-H-sol}
Suppose $F$ is a connected component of $H$ that contains no social vertices. 
Then the interpretation complexity of $\bkleq_F$ over $\Ha$ expanded by a unary predicate selecting vertices of $F$ is at most $5$.
\end{lem}
\begin{proof}
Our first goal is to interpret $\bkeq_F$, the block equivalence restricted to $F$.
To this end, we prove locality of connections within $F$ in a similar manner as in the proof of Lemma~\ref{lem:sol-to-social}.

\begin{clm}\label{cl:sol-sol-local}
For any $u,v\in F$ such that  $u,v\in G_s$ for some $s\in [n]$, there exists a path in $H$ that connects $u$ and $v$ and whose all vertices belong to blocks $G_{s-1}$, $G_s$, and $G_{s+1}$. 
\end{clm}
\begin{proof}
Among paths in $H$ that connect $u$ and $v$, choose $Q$ to be the one that minimizes the number of traversed vertices outside of blocks $G_{s-1}$, $G_s$, and $G_{s+1}$.
Since $F$ contains no social vertices, $Q$ is solitary.
In particular, by Observation~\ref{obs:local-solitary} we have that every two consecutive vertices on $Q$ belong to either to the same or to neighboring blocks.

For the sake of contradiction, suppose $Q$ traverses some vertex $r$ outside of blocks $G_{s-1}$, $G_s$, and $G_{s+1}$.
Similarly as in the proof of Lemma~\ref{lem:sol-to-social}, we choose $x$ to the earliest vertex on $Q$ before $r$ such that all vertices between $x$ (inclusive) and $r$ are outside of the block $G_s$,
and we choose $y$ to be the latest vertex on $Q$ after $y$ such that all vertices between $r$ and $y$ (inclusive) are also outside of $G_s$.
In particular, $x$ and $y$ belong to the same block, either $G_{s-1}$ or $G_{s+1}$.
Let $R$ be the infix of $Q$ between $x$ and $y$. By the same reasoning as in the proof of Lemma~\ref{lem:sol-to-social} (see Claim~\ref{cl:replacement} therein), there exists
a path $R'$ in $H$ such that:
\begin{itemize}
\item all vertices of $R'$ are solitary and belong to the same block as $x$ and $y$, which is either $G_{s-1}$ or~$G_{s+1}$;
\item $R'$ starts in some vertex $x'$ that is adjacent in $H$ to the predecessor of $x$ on $Q$;
\item $R'$ ends in some vertex $y'$ that is adjacent in $H$ to the successor of $y$ on $Q$.
\end{itemize}
Then replacing $R$ with $R'$ in $Q$ yields a path $Q'$ in $H$ that connects $u$ and $v$, but has strictly less vertices outside of blocks $G_{s-1}$, $G_s$, and $G_{s+1}$.
This contradicts the choice of~$Q$.
\cqed\end{proof}

Given $u,v\in F$, consider an \mso formula expressing the following property: $u$ and $v$ have the same modulus, and there is a path $Q$ in $H$ connecting them which traverses only vertices of
the same or neighboring moduli as $u$ and $v$. Since all paths within $F$ are solitary, by Observation~\ref{obs:local-solitary} we have that the satisfaction of this formula implies that $u$ and $v$
are in the same block. On the other hand, by Claim~\ref{cl:sol-sol-local} we have that the formula will be satisfied for all $u,v\in F$ that reside in the same block. Consequently, the presented formula,
which has quantifier rank $2$, interprets the block equivalence $\bkeq_F$.

In order to interpret the block order $\bkleq_F$, consider an \mso formula with free variables $u,v\in F$ expressing the following property.
For every path $Q$ in $H$ that leads from $u$ to $v$, if $w$ is the last vertex on $Q$ that is from the same block as $u$, then either $w=v$ or 
the successor of $w$ on $Q$ has modulus larger by one than the modulus of $v$.
To quantify $w$ we use the block equivalence $\bkeq_F$ that we interpreted in the previous paragraph, hence it is easy to obtain such a formula with quantifier rank $5$. 
To verify that this formula indeed defines the block order $\bkleq_F$, observe that if $u\in G_s$ and $w$ is the last vertex from $G_s$ on $Q$, then either $w=v$, or the successor of $w$ on $Q$
belongs to $G_{s-1}$ (which implies that $u\bkg v$) or to $G_{s+1}$ (which implies that $u\bkl v$).
\end{proof}

Lemma~\ref{lem:comps-of-H} now follows directly from combining Lemmas~\ref{lem:comps-of-H-soc} and~\ref{lem:comps-of-H-sol}: we check whether $F$ contains a social vertex, and we apply the interpretation
of Lemma~\ref{lem:comps-of-H-soc} or~\ref{lem:comps-of-H-sol} depending on the result. Hence, the proof of the Definable Order Lemma is complete.

As argued in Section~\ref{sec:idempotent}, the Definable Order Lemma implies the Idempotent Lemma. As explained in Section~\ref{sec:simon-strategy-der}, the Idempotent Lemma together with the Binary Lemma (proved in Section~\ref{sec:toolbox}) imply Theorem~\ref{thm:main} through the Simon's Lemma. Thus, we have proved Theorem~\ref{thm:main}.

\section{Conclusions}\label{sec:conclusions}

We proved that for every $k$ there is an \mso-transduction that
defines for a given graph of linear cliquewidth $k$ a width-$f(k)$
cliquewidth decomposition of this graph. A consequence of this result is
that recognizability equals \cmsoone definability for classes of graphs with bounded
linear cliquewidth. 

The obvious open question is whether our result can be generalized
from linear cliquewidth decompositions to general cliquewidth decompositions. 
The approach used in~\cite{bojanczyk2016definability} for lifting the
pathwidth case to the treewidth case heavily relies on combinatorial techniques specific to tree decompositions,
and hence it seems hard to translate the ideas to the setting of
cliquewidth decompositions.

Another question that could be asked is about the distinction between linear and general cliquewidth decompositions as outputs of decomposers. As we argued, in our main result (Theorem~\ref{thm:main}) one cannot expect that the decomposer may always output a linear cliquewidth decomposition. This is because from such a decomposition one could define a total order on the vertices of the graph, and this cannot be done for arbitrarily large edgeless graphs using a fixed \mso transduction. However, it is conceivable that under some additional assumptions on the input to the decomposer, it would be possible to transduce a linear cliquewidth decomposition. Candidates for such additional assumptions could be: (a) some stronger connectivity properties of the input graph, or (b) that the input is supplied with an arbitrary total order on the vertices (which could be used e.g. for tie-breaking).

Finally, while this might seem surprising at first glance, our proof of Theorem~\ref{thm:main} importantly uses the assumption that the input to the decomposer is an undirected graph, and not an arbitrary relational structure. This is because in our argumentation we crucially rely on the operation of flipping a graph, which is inherently tailored to the setting of undirected graphs. While appropriately defining (linear) cliquewidth for general relational structures is surprisingly non-obvious (see e.g. the discussion in Section 9.3.3 of~\cite{0030804}), there is a well-established and natural definition for directed graphs (see e.g.~\cite{CourcelleO00}). Therefore, it is natural to ask about extending  Theorem~\ref{thm:main} to general relational structures, or at least to directed~graphs.

\subsection*{Acknowledgements.} The authors thank reviewers for their careful reading and multiple insightful suggestions, especially on potential further directions presented in Section~\ref{sec:conclusions}.

\bibliographystyle{alpha}
\bibliography{bib}

\begin{thebibliography}{CMR00}

\bibitem[AK15]{adlkan15}
I.~Adler and M.M. Kant{\'e}.
\newblock Linear rank-width and linear clique-width of trees.
\newblock {\em Theor. Comput. Sci.}, 589:87--98, 2015.

\bibitem[BP16]{bojanczyk2016definability}
Miko{\l}aj Boja{\'{n}}czyk and Micha\l{} Pilipczuk.
\newblock Definability equals recognizability for graphs of bounded treewidth.
\newblock In {\em {LICS} 2016}, pages 407--416. {ACM}, 2016.

\bibitem[BP17a]{BojanczykP17}
Miko{\l}aj Boja{\'{n}}czyk and Micha\l{} Pilipczuk.
\newblock Optimizing tree decompositions in {MSO}.
\newblock In {\em {STACS 2017}}, volume~66 of {\em LIPIcs}, pages 15:1--15:13.
  Schloss Dagstuhl --- Leibniz-Zentrum f\"ur Informatik, 2017.

\bibitem[BP17b]{arxiv-BojanczykP17}
Miko\l{}aj Boja\'nczyk and Micha\l{} Pilipczuk.
\newblock Optimizing tree decompositions in {MSO}.
\newblock {\em CoRR}, abs/1701.06937, 2017.

\bibitem[CE12]{0030804}
Bruno Courcelle and Joost Engelfriet.
\newblock {\em Graph Structure and Monadic Second-Order Logic --- {A}
  Language-Theoretic Approach}, volume 138 of {\em Encyclopedia of mathematics
  and its applications}.
\newblock Cambridge University Press, 2012.

\bibitem[CMR00]{CourcelleMR00}
Bruno Courcelle, Johann~A. Makowsky, and Udi Rotics.
\newblock Linear time solvable optimization problems on graphs of bounded
  clique-width.
\newblock {\em Theory Comput. Syst.}, 33(2):125--150, 2000.

\bibitem[CO00]{CourcelleO00}
Bruno Courcelle and Stephan Olariu.
\newblock Upper bounds to the clique width of graphs.
\newblock {\em Discrete Applied Mathematics}, 101(1-3):77--114, 2000.

\bibitem[Cou90]{Courcelle90}
Bruno Courcelle.
\newblock The {M}onadic {S}econd-{O}rder logic of graphs. {I}. {R}ecognizable
  sets of finite graphs.
\newblock {\em Inf. Comput.}, 85(1):12--75, 1990.

\bibitem[GW05]{GurskiW05}
Frank Gurski and Egon Wanke.
\newblock On the relationship between {NLC}-width and linear {NLC}-width.
\newblock {\em Theor. Comput. Sci.}, 347(1-2):76--89, 2005.

\bibitem[HMP11]{hegmeipap11}
P.~Heggernes, D.~Meister, and C.~Papadopoulos.
\newblock Graphs of linear clique-width at most 3.
\newblock {\em Theor. Comput. Sci.}, 412(39):5466--5486, 2011.

\bibitem[HMP12]{hegmeipap12}
P.~Heggernes, D.~Meister, and C.~Papadopoulos.
\newblock Characterising the linear clique-width of a class of graphs by
  forbidden induced subgraphs.
\newblock {\em Discrete Applied Mathematics}, 160(6):888--901, 2012.

\bibitem[Joh98]{johansson1998clique}
{\"O}jvind Johansson.
\newblock Clique-decomposition, {NLC}-decomposition, and modular
  decomposition-relationships and results for random graphs.
\newblock In {\em Congressus Numerantium}, pages 39--60, 1998.

\bibitem[Kuf08]{Kufleitner08}
Manfred Kufleitner.
\newblock The height of factorization forests.
\newblock In {\em {MFCS} 2008}, volume 5162 of {\em LNCS}, pages 443--454.
  Springer, 2008.

\bibitem[OS06]{OumS06}
Sang{-}il Oum and Paul~D. Seymour.
\newblock Approximating clique-width and branch-width.
\newblock {\em J. Comb. Theory, Ser. {B}}, 96(4):514--528, 2006.

\bibitem[Oum05]{Oum05}
Sang{-}il Oum.
\newblock Rank-width and vertex-minors.
\newblock {\em J. Comb. Theory, Ser. {B}}, 95(1):79--100, 2005.

\bibitem[Sim90]{Simon90}
Imre Simon.
\newblock Factorization forests of finite height.
\newblock {\em Theor. Comput. Sci.}, 72(1):65--94, 1990.

\bibitem[Wan94]{wanke1994k}
Egon Wanke.
\newblock {$k$-NLC} graphs and polynomial algorithms.
\newblock {\em Discrete Applied Mathematics}, 54(2-3):251--266, 1994.

\end{thebibliography}

\end{document}